\documentclass[epj,numbook]{svjour}

\usepackage{graphics}
\usepackage{graphicx}
\usepackage{amsmath}
\usepackage{subfigure}
\usepackage{anysize}
\usepackage{amssymb}
\usepackage{dcolumn}
\usepackage{multirow}
\usepackage{longtable}

\begin{document}

\title{Vortex Washboard Voltage Noise in Type-II Superconductors}
\author{Thomas J. Bullard\inst{1,3}, Jayajit Das\inst{2}, 
        George L. Daquila\inst{1}, and Uwe C. T\"auber\inst{1}}
\authorrunning{T.J. Bullard et al.}
\institute{Department of Physics and Center for Stochastic Processes and 
  Engineering, Virginia Polytechnic Institute and State University, 
  Blacksburg, VA 24061-0435, USA \and
  Department of Chemical Engineering, 
  Massachusetts Institute of Technology, Cambridge, MA 02139, USA \and
  U.S. Air Force, Wright--Patterson Air Force Base, OH 45433-5648, USA}

\date{\today}

\abstract{\rm
In order to characterize flux flow through disordered type-II superconductors, 
we investigate the effects of columnar and point defects on the vortex velocity
/ voltage power spectrum in the driven non-equilibrium steady state. 
We employ three-dimensional Metropolis Monte Carlo simulations to measure 
relevant physical observables including the force-velocity / current-voltage 
(I-V) characteristics, vortex spatial arrangement and structure factor, and 
mean flux line radius of gyration.
Our simulation results compare well to earlier findings and physical intuition.
We focus specifically on the voltage noise power spectra in conjunction with 
the vortex structure factor in the presence of weak columnar and point pinning 
centers.  
We investigate the vortex washboard noise peak and associated higher harmonics,
and show that the intensity ratios of the washboard harmonics are determined by
the strength of the material defects rather than the type of pins present.
Through varying columnar defect lengths and pinning strengths as well as 
magnetic flux density we further explore the effect of the material defects on 
vortex transport.
It is demonstrated that the radius of gyration displays quantitatively unique
features that depend characteristically on the type of material defects present
in the sample.
}

\maketitle

\section{Introduction}

The physics of interacting vortex lines in high-tem\-perature superconductors 
subject to strong thermal fluctuations and pointlike or extended disorder is 
amazingly rich and has been a major research focus in condensed matter physics 
in the past two decades \cite{BLAT}.
Vortex motion plays a crucial role in the transport properties of type-II 
superconductors in external magnetic fields. 
In the presence of a sufficiently large applied current magnetic flux lines 
will experience a Lorentz force and drift perpendicular to both the current and
applied magnetic field.  
The motion of these current-encircled magnetic flux filaments induces an 
electric field parallel to the applied current resulting in power dissipation 
and hence an Ohmic voltage drop across the material proportional to the 
velocity of the vortices.  
While an externally applied current serves as a driving force, underlying 
defects in the superconducting material inhibit vortex motion.
Material defects pin vortices below a critical applied force, and also play a 
significant role in their motion above that threshold, providing, on a 
coarse-grained description level, an effective friction or viscosity.  

Aside from the obvious relevance for technological applications, driven 
magnetic flux lines in type-II superconductors also represent one of the few 
cleanly experimentally realizable systems of interacting particles in a 
non-trivial non-equilibrium steady state.  
A thorough understanding of the ensuing phase diagram and full characterization
of each emerging steady state should shed light on the rich and still rather 
incompletely understood features of non-equilibrium systems in general.
Since stochastic fluctuations and intrinsic correlations typically play a 
significant role away from thermal equilibrium, it is desirable for each 
emerging stationary state to attain a thorough quantitative understanding of 
the fluctuations.

In superconductors specifically, pinning effects on vortex motion are also 
reflected in the voltage noise power spectrum \cite{CLEM}.  
For instance, slightly abo\-ve the critical force, in the presence of strong 
disorder, depinning of vortices is observed to proceed via flow through plastic
channels \cite{BRANDT}.  
These `rivers' of vortices form at different locations in the sample, and flow 
around `islands' of temporarily trapped flux lines resulting in incoherent 
motion.  
Such behavior has been observed experimentally \cite{MATS1}, as well as in 
two-dimensional computer simulations \cite{JENS1}, and is characterized in the 
velocity or voltage frequency power spectrum by a broadband noise signal which 
obeys a $1/\omega^\alpha$ power law, as also demonstrated in recent 
three-dimensional numerical work \cite{VEST}.

Well above the critical force it has been observed that vortices are more 
translationally ordered than at low velocities \cite{YARIN}.  
It might be expected that at a sufficiently high drive, the vortices would form
a moving Abrikosov lattice since the effective pinning force from the disorder 
on each vortex varies rapidly and would therefore be less effective 
\cite{KOSH}.  
However, it has been shown by Le~Doussal and Giamarchi \cite{DOUS1} that some 
modes of the disorder are not affected by the motion even at large velocities.
As a result the vortices enter what has been termed a `moving glass' phase.  
Here, subtle competitions between elastic energy, disorder, and dissipation 
lead to the transverse displacements becoming pinned into preferred 
time-independent configurations resulting in stationary two-dimensional 
channels.
In the moving glass phase, vortices thus follow each other in a manner similar 
to beads on a wire.

There exist a few possible coupling regimes between these elastic channels.  
For strong point disorder approximately parallel elastic channels are 
completely decoupled, while the periodicity along the direction transverse to 
the drive is maintained.  
Here sharp delta-function Bragg peaks with nonzero reci\-procal-lattice vector 
components along the direction of motion are lost, while peaks with only 
transverse components remain.  
This regime is known as the `moving transverse glass' or `moving smectic' 
phase \cite{DOUS1}, and is supported by recent numerical simulations in two and
three dimensions \cite{GOTCHA,CHEN}.  
On the other hand, for weak point disorder, or large velocities, relative 
deformations grow only logarithmically with distance; hence, the vortex 
structure maintains quasi long-range order corresponding to complete elastic 
coupling between vortices.  
This state is known as a `moving Bragg glass' and is characterized by 
algebraically divergent structure factor peaks at small reciprocal lattice 
vectors \cite{DOUS1}.

The interaction of the moving Bragg glass with the underlying material defects 
is manifested by a characteristic peak in the power spectrum corresponding to 
the periodicity of the vortex lattice \cite{DOUS1}.  
Defects in the material temporarily slow vortices resulting in `stick-slip' 
motion.  
In a structurally ordered phase such as the moving Bragg glass, this behavior 
is repeated, resulting in a periodically varying average overall velocity.  
With the lattice vector of the Bragg glass oriented in the flow direction, the 
resulting characteristic frequency associated with this motion is known as the 
`washboard' frequency.  
Related phenomena are certainly not unique to dri\-ven flux lines in type-II 
superconductors, but are, for example, well-established in charge- and 
spin-density wave systems \cite{GRUNER}.

Random point disorder need not be the only pinning structure in superconducting
materials.  
Col\-umnar disorder may also be introduced into materials for the purpose of 
increasing the critical current \cite{CIVALE}.  
Far above the depinning threshold a `moving Bose glass' is formed in the 
presence of columnar pins \cite{DOUS1}.  
Owing to the correlated nature of the disorder along the length of each vortex,
the structure function tends to resemble that of the moving transverse glass.  
Furthermore, akin to the equilibrium Bose glass (i.e., the disorder-dominated 
amorphous structure formed by flux lines localized by correlated defects 
\cite{NEL1}), the moving Bose glass displays a diverging tilt modulus 
\cite{DOUS1,OLIVE}. 
Whether a unique voltage noise signal exists for this phase (as well as for the
moving transverse glass) is unclear, and subject to this present investigation.

% Early experiments on washboard noise in superconductors date back to Fiory 
% \cite{FIORY} who, by applying both an AC and DC current to an aluminum film 
% in the mixed state, observed a drop in resistance when the time for the 
% Abrikosov lattice to travel a distance of one lattice constant (i.e., the 
% washboard period) corresponded to integer or half-integer multiples of the AC
% current period. Similar results were obtained by Harris {\em et al.} 
% \cite{HARRIS} for high-$T_c$ materials.
%
% Compelling evidence for washboard motion has been obtained in the creep 
% regime in various experiments. Troyanovsky {\em et al.} \cite{TROY} observed 
% washboard motion in NbSe$_2$ crystals using scanning tunneling microscopy.  
% Images of vortex positions were taken at a sampling frequency high enough to 
% observe their motion in the creep and plastic regimes by compiling a sequence
% of images into a movie. For columnar defects plastic motion was recorded for 
% the highest applied magnetic fields. For point defects a moving vortex 
% lattice was observed for a number of different applied magnetic fields. By 
% tracking the velocity of a number of individual vortices a frequency 
% corresponding to the washboard frequency was measured.

The washboard frequency has been observed in a number of experiments
\cite{FIORY,HARRIS,TROY}.
However, direct observation of the washboard noise was achieved by Togawa 
{\em et al.} \cite{TOGAWA}, who obtained voltage noise spectra of BSCCO 
crystals in the mixed state subject to a constant current for various applied 
magnetic field strengths.  
For low magnetic fields broadband noise was observed and attributed to plastic 
vortex flow.  
As the applied magnetic field was increased, the broadband noise signal reached
a maximum value and then decreased again, while a narrowband noise peak that
corresponded to the washboard frequency emerged.  
Upon increasing the magnetic field further, the characteristic frequency of the
narrowband noise grew owing to a tighter flux line packing of the vortices, and
hence a shorter vortex lattice constant.  
The narrowband noise signal also decreased in height and increased in width.  
The reason for this is apparently still not fully understood.

Washboard noise has also been detected unambiguously in a number of 
two-dimensional numerical simulations; as characteristic examples, we mention
the following:
Olson {\em et al.} \cite{OLSON1} performed mole\-cular dynamics simulations of 
vortices that were dri\-ven through a system of randomly placed defects.  
Upon varying the drive strength they noticed that the number of regimes 
available to the system above the plastic flow phase depended on the vortex 
interaction strength.  
For systems with intermediate to high interaction strength and large applied 
drive, the vortices entered a coupled channel regime indicated by sixfold 
coordination of the structure.  
In this regime washboard noise was observed.  
Furthermore, it was noted that the washboard signal intensity decreased as the 
system size was increased.  
It is believed that this was due to multiple domains forming in the vortex 
lattice resulting in decoherence of the noise signal.

Kolton {\em et al.} \cite{KOLTON1} also performed two-dimension\-al molecular 
dynamics simulations to investigate Fiory steps \cite{FIORY} for different 
vortex velocities.  
In order to understand the relationship between these steps and the temporal 
order in the different regi\-mes, the voltage power spectrum was investigated 
without an applied ac drive.  
The authors observed the evolution of the power spectrum from broadband noise 
to the emergence of a narrowband peak as the applied dc drive was increased, 
and also found higher harmonics.  
Two-dimensional computer simulations employing Langevin dynamics and varying 
drive and pinning strength have been performed as well \cite{FANG}, with 
similar washboard noise results.

In contrast, we are aware of only one three-dimensional investigation of the 
washboard noise.  
Using an anisotropic XY model Chen and Hu \cite{CHEN} investigated the 
first-order transition from the moving Bragg glass to the moving smectic.  
In the moving Bragg glass phase narrowband noise that corresponded to the 
washboard frequency was observed along with harmonics for various driving 
strengths at zero temperature.  
The washboard peak itself persisted for $T>0$, while its harmonics disappeared.
The authors also found that the moving Bragg glass turned into a moving liquid 
at high drive, and hence that the washboard signal was destroyed.  
They argue this to be due to thermally activated vortex loops inducing 
dislocations in the Bragg glass \cite{CHEN}.

The purpose of this work \cite{TOM} is threefold.  
First, we wish to establish confidence in our novel simulation approach to 
modeling vortex motion by qualitatively comparing our results to well-known 
superconducting vortex behavior.  
Second, we wish to investigate the evolution of the narrowband noise associated
with the washboard frequency for increasing vortex density (i.e., increasing 
magnetic flux density) in the presence of randomly distributed point as well as
columnar defects.  
Finally, we are interested in studying the effects of different types of 
pinning centers on the narrowband voltage noise.  
Below the critical current, in the presence of columnar defects, it has been 
predicted that vortices hop between pinning sites temporarily trading elastic 
deformation energy in the form of double-kinks and half-loops for a lower 
overall energy configuration \cite{NEL1}.  
Well above the depinning current remnants of these half-loop excitations and 
double kinks may still exist.  
At high driving values these remnant excitations would occur predominantly in 
the direction of the drive, while other excitations would be suppressed by the 
localizing effect of the columnar pins.  
Obviously, the impact of such vortex excitations on the power spectrum cannot 
be addressed by a two-dimensional simulation, but require a full 
three-dimensional model.

Based on the effective free energy for interacting magnetic flux lines in the
London approximation, and subject to attractive pinning centers \cite{NEL1}, we
have developed a three-dimensional Monte Carlo simulation code \cite{DAS} to 
study the effects of disorder on the velocity / voltage power spectrum and the 
two-dimensional static vortex structure factor in the plane transverse to the 
magnetic field for driven vortices in the non-equilibrium steady state. 
Specifically, we compare results for point and columnar defects.  
We also measure the average radius of gyration in order to examine the effects 
of the different defect types on the shape or thermal `wandering' of the 
elastic flux lines along the magnetic field ($z$) direction. 
The simulation results reported here should be contrasted with our earlier 
findings for non-interacting flux lines in the presence of various disorder 
distributions \cite{DAS}.

Our results display many similar features for both defect types.  
As the vortex density is increased for systems with either weak point or 
correlated disorder, positional ordering is observed to increase in the 
structure factor plot.  
For columnar defects the vortex structure factor is found to change from that
characteristic of a typical liquid, to a smectic, and eventually an ordered
triangular lattice.  
For the case of point defects, we only observe the triangular array in the 
parameter region studied here.  
We find that the structure factor plots at low vortex densities in the presence
of point disorder appear qualitatively similar to the results for columnar 
defects at higher densities.  
As the structure factor begins to display positional order in the direction of 
the drive a narrowband noise signal in the velocity noise power spectrum 
corresponding to the washboard effect is detected for both defect types.  
Associated with the washboard peak are harmonics, the ratios of
which initially appear to be related to the type of pinning defect in the
system.  
We present results that suggest these harmonic ratios are in fact not dependent
on the type of pinning centers present in the sample.

To further examine these potentially distinct effects on the velocity or 
voltage power spectrum, we vary the effectiveness of the pinning centers.  
For a fixed vortex density the positional arrangement of the pinning sites in 
the system is changed from randomly distributed (i.e., point-like defects) to 
correlated along the $z$ axis (i.e., columnar defects).  
The velocity fluctuation power spectra as well as the structure factor plots 
are examined for this series of simulations as well.  
We then compare our findings to the results obtained with increasing point 
defect pinning strength.  
We find that whereas the power spectrum and structure factor evolve in a 
qualitatively similar manner when varying the `pinning effectiveness' through 
either method, there appear marked differences in the behavior of the mean 
radius of gyration.  
Namely, as the point pinning strength increases, vortices tend to stretch and 
deform following a $r_g \propto U^2$ behavior, where $U$ represents the 
pinning potential depth.  
In contrast, we observe the radius of gyration to saturate for increasing 
columnar defect length.  
The behavior is best described phenomenologically as $r_g \propto e^{-l_0/l}$ 
where $l$ denotes the length of the columnar defects (and $l_0$ gives the
length scale in $z$ direction).

In the following section 2, we describe our model and the simulation algorithm 
in detail.  
Section 3 contains our simulation results, as already summarized above.  
Finally, in section 4, we conclude and provide an outlook for further 
investigations.

\section{Model Description and Simulation Algorithm}

In our Monte Carlo simulations, vortices are considered in the London 
approximation (with the London penetration depth large compared to the 
coherence length, $\lambda \gg \xi$) as discretized elastic lines \cite{NEL1}
(see also Refs.~\cite{ROSSO,SEN}).  
The elastic energy associated with the line tension of $N_v$ flux lines is 
taken to be
\begin{equation}
  E_L = \frac{\epsilon_1}{2} \sum_{i=1}^{N_v} \int_{0}^{L} dz
  \bigg\arrowvert\frac{d\boldsymbol r_i(z)}{dz}\bigg\arrowvert^2 ,
\label{etension}
\end{equation}
where $\boldsymbol{r}_i(z)$ describes the configuration of the $i$th vortex by
specifying its two-dimensional position $\boldsymbol{r}$ as function of the
coordinate $z$ ($0 \leq z \leq L$) along the magnetic field direction.  
The line stiffness is given by $\epsilon_{1} = \epsilon_{0} \ln \bigl(
\xi_{ab} / \lambda_{ab} \bigr) \Gamma^{-2}$, where $\lambda_{ab}$ is the 
in-plane London penetration depth, and $\xi_{ab}$ the in-plane superconducting 
coherence length.  
$\epsilon_{0} = \bigl( \phi_{0} / 4\pi\lambda_{ab} \bigr)^2$ sets the overall 
energy scale, and $\phi_{0} = hc / 2e$ is the magnetic flux quantum.  
The expression (\ref{etension}) for the elastic energy holds if 
$\big\vert d\boldsymbol{r}_i(z) / dz \big\vert^2 \ll \Gamma^{-1}$, where 
$\Gamma^2 = M_{z}/ M_\perp$ denotes the effective mass ratio for the elastic 
line.  
In this study we model high-$T_{c}$ materials for which $\Gamma \gg 1$.  
In the simulation each flux line is represented by $N_{p}$ points located at 
$(\boldsymbol{r}_{i},z_{i})$.  
Each point is confined to a constant $z_{i}$ (a separate $ab$ plane) and 
interacts with its nearest neighbors above and below via a simple harmonic 
potential.

The total interaction energy between all pairs of distinct vortices is
\begin{equation}
  E_{\rm int} = \sum_{i \not= j}^{N_v} \int_{0}^{L} 
  V\Bigl(|\boldsymbol{r}_{i}(z)-\boldsymbol{r}_{j} (z)|\Bigr) \, dz \, ,
\end{equation}
with the pair potential $V(r) = 2\epsilon_{0} K_{0}(r / \lambda_{ab})$.
Here, $K_{0}$ is the modified Bessel function of zeroth order, and can be 
described qualitatively as diverging logarithmically as $r\rightarrow 0$ and 
decreasing exponentially for long distances $r \gg \lambda_{ab}$.  
Interactions between vortices occur only within the $ab$ planes (i.e., 
consistent with the London limit we neglect any cross-plane interactions); this
approximation is valid as long as the requirements for Eq.~(\ref{etension}) are
satisfied.  
For the simulation, we consider a system of extension $L_x$ and $L_y$ in the 
$x$ and $y$ directions with periodic boundary conditions.
For each vortex pair, we compute its minimal distance within this rectangle and
its periodic images adjacent to it, and use that distance to evaluate the
interaction potential. 
As a consequence of this nearest-image approximation, the interaction potential
is cut off at distance min$(L_x/2,L_y/2)$.
To minimize the effects of the cut-off, $\lambda_{ab}$ has been decreased to 
prevent numerical artifacts observed in the simulation \cite{fnote1}.
We have also run Monte Carlo simulations for a cutoff length twice the original
length used in the bulk of this study.
To accommodate the increase in length the system area was increased by a factor
of four.  
For the larger system the washboard noise is recovered for both columnar and 
point defects, and higher harmonics are observed for point defects.

Material defects in the system are represented by a distribution of cylindrical
potential wells, 
\begin{equation}
  E_D = \sum_{j=1}^{N_v} \int_{0}^{L} 
  V_D\Bigl(\boldsymbol{r}_j(z)\Bigr) \, dz \, ,
\end{equation}
with $V_D\Bigl(\boldsymbol{r}_j(z)\Bigr) = \sum_{k=1}^{N_D} U 
\Theta\Bigl(b_0-|\boldsymbol{r}_j(z)-\boldsymbol{r}_k^{(p)}|\Bigr)$.
Here $b_0$ is the pin radius in the $ab$ plane, $\Theta$ denotes the Heaviside 
step function, $\boldsymbol{r}_k^{(p)}$ indicates the location of the $k$th 
pinning center, and $U$ characterizes the well depth.

Finally, in the presence of an external current the flux lines experience a 
force per unit length 
$\boldsymbol{f}_L = \phi_{0} \hat{z} \times \boldsymbol{J} / c$,
therefore a corresponding work term is introduced:
\begin{equation}
  W = -\sum_{i=1}^{N_v} \int_{0}^{L} \boldsymbol{f}_L \cdot 
  \boldsymbol{r}_{i}(z) \, dz \, .
\end{equation}
This work contribution favors vortex motion along the direction of the force 
while suppressing motion against it.  
For this investigation, the force is always applied in the $x$ direction.  
The total energy of the system reads 
$E_{\rm tot} = E_L + E_{\rm int}+ E_D + W$.

In our simulations the applied magnetic field is taken parallel to the $c$ axis
(oriented parallel to $\hat{z}$); therefore, at time $t=0$ straight lines are 
placed vertically in a system of size $L_x \times L_y \times L_z$ with periodic
boundary conditions in all directions.  
We have found that the initial configurations of the individual lines do not 
affect the steady state attained at long times.  
Defect centers at positions $\boldsymbol{r}_k^{(p)}$ are also distributed 
throughout the system, either randomly or aligned parallel to the $c$ axis to 
model columnar pins. 
The state of the system is then updated according to standard Monte Carlo 
Metropolis rates.  
When the number of attempted updates is equal to the number of points that make
up a flux line multiplied by the total number of lines in the system, this 
constitutes a single Monte Carlo step (MCS) and serves as the unit of time in 
the simulation.  
While this `driven diffusive system' simulation approach, introduced by Katz, 
Lebowitz, and Spohn \cite{KATZ}, is likely best suited to model thermally 
activated motion close to equilibrium, we find that our results in the driven 
regime, considerably away from equilibrium, are quite similar to those observed
in other studies, as mentioned above.

Nevertheless, it is not at all obvious even for steady states far from thermal 
equilibrium which choices of `microscopic' Monte Carlo rates yield the most 
`realistic' representations of an experimental system, even in a coarse-grained
view.  
Gotcheva {\em et al.} \cite{GOTCHA} have recently brought into question the 
general validity of the driven diffusive Metropolis Monte Carlo dynamical 
method in their simulations for vortices on a two-dimensional lattice.  
The non-equilibrium steady states obtained using Metropolis Monte Carlo and 
continuous-time Monte Carlo dynamical rules, respectively, were compared as a 
function of temperature and driving force.  
The results differed dramatically depending on which dynamical rule was used.  
In some instances at least, the Metropolis algorithm yielded a spatially 
disordered moving steady state while the continuous-time Monte Carlo rules 
preserved positional order (in finite-size systems) over much of the 
non-equili\-brium phase diagram.  
The authors argue convincingly that the lack of order observed in the 
Metro\-polis simulation was due to intrinsic randomness in the updating rules. 
It would certainly be interesting and worthwhile to probe to what extent these 
findings also apply to three-dimensional off-lattice simulations, which are
presumably less likely to be stuck in long-lived metastable configurations.  
In our current study we find that spatial order survives in the non-equilibrium
steady state for an extended range of flux density values.

Quite generally, it is crucial for the analysis of out-of-equilibrium systems 
to carefully investigate alternative approaches to the description of their 
dynamics in order to probe their actual physical properties rather than 
spurious artifacts inherent in any mathematical modeling.
Different mathematical and numerical representations of non-equilibri\-um 
systems in fact rely on various underlying {\em a-priori} assumptions that must
be tested {\em a-posteriori} by comparing their respective results.
For example, many computer simulation studies of driven flux line systems 
invoke stochastic Langevin equations, wherein one assumes that all fast degrees
of freedom are aptly captured in terms of uncorrelated white noise (see, e.g.,
Refs.~\cite{OLSON1,OLSON2,FANG}). 
Such a mesoscopic representation of the dynamics is usually adequate in thermal
equilibrium where the form of the input noise correlations is severely 
restricted by fluctua\-tion-dissipation relations.
Yet the large-scale and long-time properties of nonlinear Langevin stochastic 
differential equations away from thermal equilibrium are well-known to be often
drastically affected by the functional form or even strength of the assumed 
noise correlations, which are not uniquely determined by Einstein relations any
more.
One would, e.g., suspect that the flux line correlations along the magnetic 
field axis should be reflected in the noise spectrum and relaxational features.
It is therefore imperative to test a variety of different numerical methods and
compare the ensuing results in order to identify those properties that are 
generic to the physical system under investigation.

The parameter values used in our simulations correspond to typical high-$T_c$ 
materials, and are reported in units of $b_0$ (the pin radius) and interaction 
energy scale $\epsilon_0$.  
The parameters $\xi_{ab}$, $\epsilon_1$, and $\Gamma^2$ are chosen to be 
$0.5 \, b_0$, $0.25 \, \epsilon_0$, and $16$, respectively.  
In this study we examine vortex motion in the weak pinning regime.  
The pinning potential $U$ has been given a value of 
$U_0 = 0.03125 \, \epsilon_0$ while the predicted value is approximately 
$0.5 \, \epsilon_0$ \cite{NEL1}.   
The penetration depth $\lambda_{ab}$ is assigned a value of $16 \, b_0$ which 
is about $1/3$ of typical high-$T_c$ values.   
As previously mentioned this choice was made in order to minimize artifacts due
to the interaction cut-off.  
The average separation distance between randomly distributed defects in each 
$ab$ plane is taken to be $15 \, b_0$.  
The maximum distance for a point on any flux line to move is limited to 
$b_0 / 2$, to help avoid vortices `hopping' over defects.
We have used a discretization along the field direction of $L_z = 20$ parallel 
planes.

As is well-known in finite-size vortex simulations, a square planar geometry 
favors ordering into a square lattice whose configurational energy is only 
slightly above that of the triangular Abrikosov lattice \cite{KLEIN}.
We have thus chosen the system's planar aspect ratio as $L_x:L_y = 2:\sqrt3$ in
order to easily accommodate a triangular lattice.
The vortices are then placed in the system prearranged in a triangular array 
with a lattice vector oriented along the x-direction. 
Choosing this aspect ratio allows an even square number of vortices to `fit' 
in the system while arranged in this configuration.
In the simulation runs, the vortex lattice arranged in local energy minima 
configurations by either aligning a lattice vector parallel to the system's 
horizontal axis, rotating the lattice vector by $30^{\circ}$ from the 
horizontal, or by arranging such that the lattice twisted about the system at a
chiral angle satisfying periodic boundary conditions.   
When in our simulations an external force is applied and the vortex system
driven, we find that the lattice maintains its initial orientation.  
However, by approximately doubling the defect pinning strength the lattice 
reorients such that the principal lattice vector points in the direction of the
applied drive.  
Drive-induced reorientation has been observed in experiments \cite{SCHMID} and 
simulations \cite{FANG}.  

Since the present study is primarily concerned with the effect of defect 
correlations on the dynamics, the temperature is chosen such that 
$T / T^* <1$, where $T^* = k_{B}^{-1} \sqrt{\epsilon_1 U_0} b_0$ is the 
temperature above which entropic corrections due to thermal fluctuations become
relevant for pinned flux lines \cite{NEL1}.
Here, thermally induced stretching and wandering of the flux lines are largely 
suppressed (as long as the vortices remain pinned), and the results can be 
interpreted in terms of low-temperature kinetics.
Our simulations are thus usually run at $k_{\rm B} T$ per unit length equal to 
$0.004 \, \epsilon_0$.  
The average or center of mass (CM) velocity for each vortex is then calculated 
and averaged over all vortices:
\begin{equation}
  \boldsymbol{v}_{\rm cm} = \frac{1}{N_v} \sum_{i=1}^{N_v}
  \frac{\boldsymbol{r}_{\rm cm_{i}}(t+\tau) 
  - \boldsymbol{r}_{\rm cm{_i}}(t)}{\tau} \, ,
\end{equation}
where $\boldsymbol{r}_{\rm cm{_i}}$ denotes the center-of-mass position of the 
$i$th vortex, and $\tau$ is the time interval between measurements.  
$\tau$ is set to $30$ MCS, and the simulation is then run for $10^5$ MCS to 
arrive at a steady state.  
Data are subsequently collected for the next $2.5 \times 10^5$ MCS.

From the collected data, we obtain the two-dim\-ensional static structure 
factor in the plane transverse to the magnetic field, 
\begin{equation}
  S({\bf k}) = \int \langle \rho({\bf 0}) \rho({\bf r}) \rangle \,
  e^{- i {\bf k} \cdot {\bf r}} \, d{\bf r} \, ,
\end{equation}
where $\langle \rho({\bf 0}) \rho({\bf r}) \rangle$ denotes the density-density
correlation function, for the driven vortices in the non-equilibrium steady 
state. 
We also measure the average radius of gyration, 
\begin{equation}
  r_g = \left(\frac{1}{N_v L} \sum_{i=1}^{N_v} \int_0^L 
  [r_{{\rm cm}_i} - r_i(z)]^2 \, dz \right)^{1/2} , 
\end{equation}
in order to examine the effects of the different defect types on the shape or 
thermal `wandering' of the elastic flux lines along the magnetic field ($z$)
direction. 
Additional information about the detailed dynamics is encoded in the velocity 
fluctuation power spectra 
\begin{equation}
  S(\omega) = \left| \int v(t) \, e^{i \omega t} \, dt \right|^2 .
\end{equation}
These velocity power spectra have been appropriately windowed in order to 
minimize spectral leakage.
Since vortex motion across the sample induces a voltage drop, the velocity
fluctuations are experimentally directly accessible as the measured voltage
noise.
Each of the above observables is measured for various vortex densities and 
defect configurations.  
Vortex densities are reported as a count of the number of vortices in a unit 
system of area $A = 150 \frac{2}{\sqrt{3}} \, b_0 \times 150 \, b_0$.  
The vortex structure factor and velocity noise plots are averaged over time and
typically over $40$-$50$ disorder realizations.

\section{Monte Carlo Simulation Results}

\subsection{Current-Voltage (I-V) Characteristics}

In order to validate our model and simulation algorithm we have examined the 
average vortex velocity as a function of the applied force and compared our
results to well-established experimental and numerical findings.  
In experiments, the driving force is proportional to the externally applied 
current, and the induced voltage across the sample proportional to the mean 
flux line velocity, hence the current-voltage (I-V) characteristics are given 
by the force-velocity curves in our simulations.

\begin{figure}
\begin{center}
  \subfigure{\label{randomiv}\includegraphics[scale=.6]{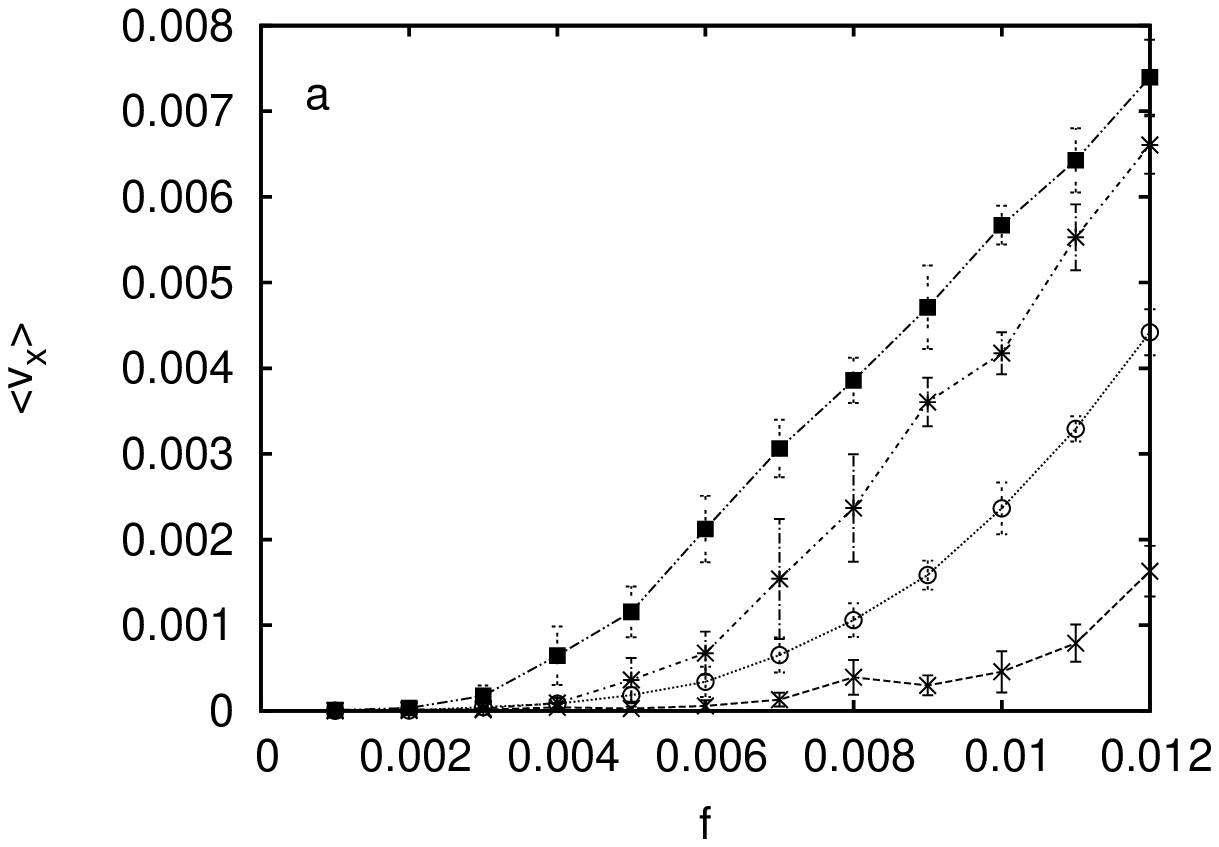}}
  \vskip -0.2 truecm
  \subfigure{\label{pointsiv}\includegraphics[scale=.6]{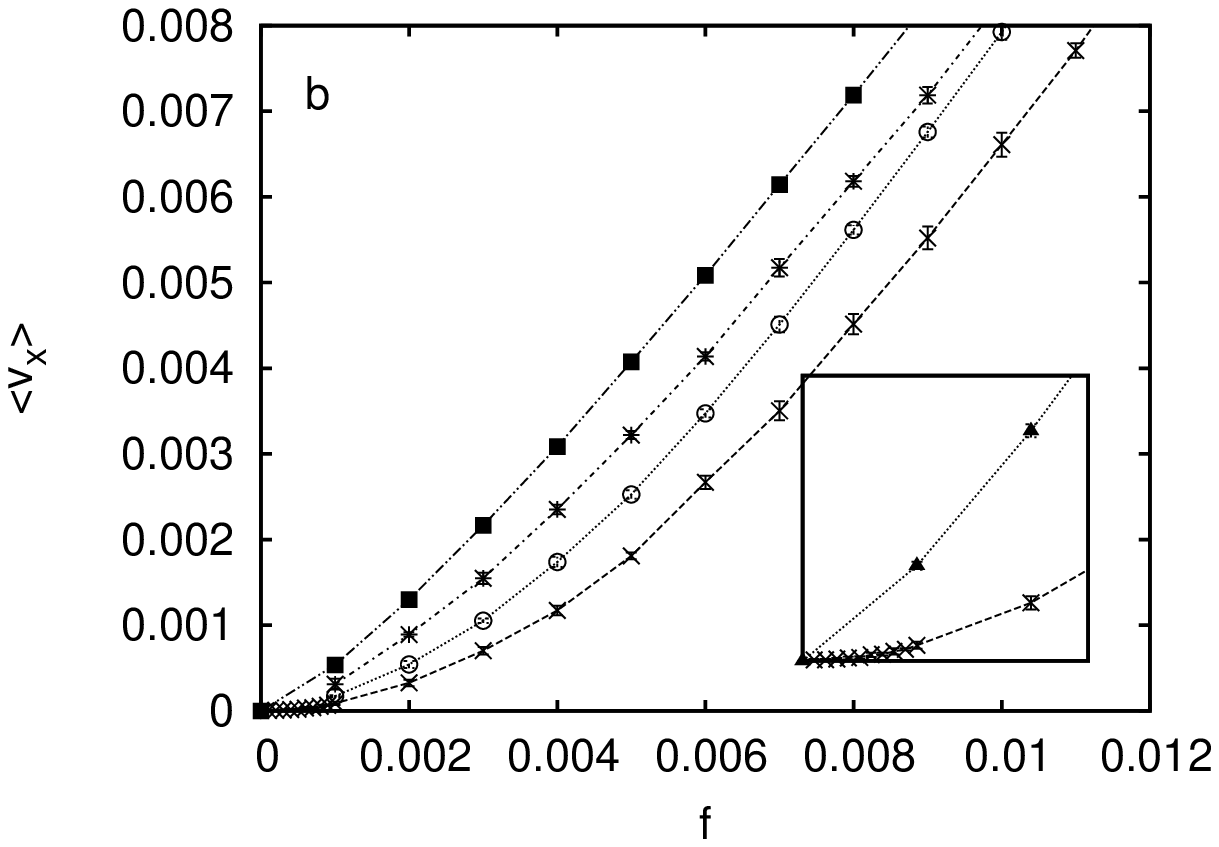}}
  \vskip -0.2 truecm
  \subfigure{\label{depinthr}\includegraphics[scale=.6]{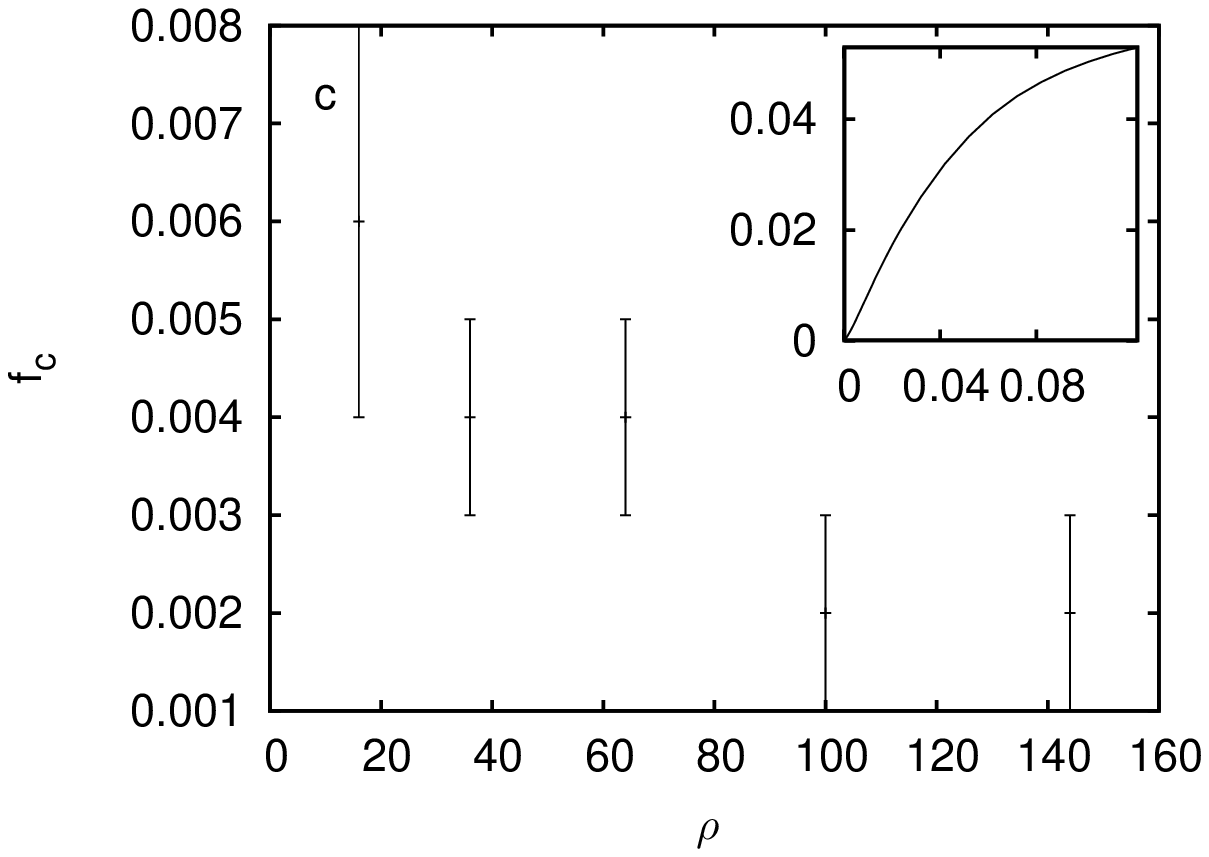}}
\end{center} 
  \vskip -0.3 truecm
\caption{Velocity (induced voltage) vs. force (applied current) curves 
  (I-V characteristics) for (a) randomly distributed columnar pins, and 
  (b) point defects.  
  For both defect types the depinning threshold decreases with increasing 
  vortex / flux density.  
  This can be seen most clearly for point defects in the inset in (b), which
  amplifies the low current regime: here, the force values on the $x$ axis of 
  the inset range from $0$ to $0.0025$, the velocity ranges from $0$ to 
  $0.0016$ and is given in units of $b_0 /$ MCS (pin radius per Monte Carlo
  step).  
  The following symbols represent vortex densities reported as the areal 
  density $\rho$ in a fixed area 
  $A = 150 \frac{2}{\sqrt3}\, b_0 \times 150\, b_0$: 
  $\blacktriangle - 144 / A = 0.00554/b_0^2$, 
  $\blacksquare - 100 / A = 0.00385/b_0^2$, $* - 64 / A = 0.00246/b_0^2$, 
  $\bigcirc - 36 / A = 0.00139/b_0^2$, $\times - 16 / A = 0.00062/b_0^2$.  
  (c) Estimate of the depinning force $f_c$ as function of the vortex density
  $\rho$ for random columnar defects.
  The inset depicts how the I-V curve for $\rho = 144 / A$ saturates at high 
  drive values. 
  In all plots, the data points are connected as a guide for the eye.}
\label{ivresults}
\end{figure}

Our simulation results for systems with random\-ly distributed columnar pins 
and point defects are plotted for various flux densities in 
Fig.~\ref{ivresults}.  
In addition to averaging over all flux lines in the system, the data are 
averaged over time (25,000 Monte Carlo steps) and five disorder realizations.
For both graphs the number of effective pinning sites is equal, as are the
pinning strengths (per unit length) $U_0$ for all the defects.

The vortices are seen to remain pinned for both columnar and point defects up 
to a critical depinning force, beyond which the system takes on an average 
velocity that increases with increasing applied drive.  
As expected the depinning threshold is higher for columnar defects 
[Fig.~\ref{randomiv}] since the attractive force from each pinning site adds 
coherently over the length of the flux line.  
The inset in Fig.~\ref{pointsiv} depicts the region of the curve associated 
with depinning for point defects, showing just the data for the densest and 
most dilute systems.  
We observe (again as expected) that denser vortices depin at lower driving
currents for either type of pinning center; estimates for the depinning current
for columnar defects as obtained from our data are depicted in 
Fig.~\ref{depinthr}. 
As the repulsive interactions between neighboring vortices grow with flux 
density, this helps the vortices to overcome the defect pinning potentials.
We find that these results are qualitatively comparable to experimental 
findings in high-temperature superconducting materials (see, e.g., 
Refs.~\cite{ANDO,AMMOR,QIANG}), and in accord with results obtained by means of
Langevin molecular dynamics simulations \cite{OLSON3} (in the three-dimensional
regime, see also Ref.~\cite{KOLTON2}), giving us confidence in the validity of
our model and algorithm.

However, we note two artifacts in the I-V results occuring at higher driving
current and large vortex densities.  
First, while at large currents linear Ohmic behavior should ensue, we find that
the velocity values saturate at high applied drive values, see inset of 
Fig.~\ref{depinthr}.  
This is due to a maximal step size limitation imposed to avoid vortices 
'hopping' over pinning defects in the system.  
Second, the I-V curves are observed to cross at larger densities due to the 
fact that the system is updated locally rather than globally, in the Metropolis
algorithm.  
At higher vortex densities larger local moves are suppressed by the repulsive 
potential of nearest neighbors resulting in a lower average velocity and an I-V
curve with a lower slope than that of lower density systems.

% However, our numerical algorithm ceases to pro\-duce physically reliable I-V
% results at high driving currents:
% The velocity versus force curves generated by our simulations are observed to
% cross at higher drive values. 
% This is especially clear for the point defect data [Fig.~\ref{pointsiv}].  
% Furthermore, while at large currents linear Ohmic behavior (i.e., overdamped 
% motion) should ensue, we find that the velocity values saturate at high 
% applied drives in the inset of Fig.~\ref{depinthr}.  
% This artifact is due to the following limitation present in the simulation 
% algorithm:  
% In order to guarantee interaction of the vortices with defects the maximal 
% `hopping' distance for the vortices has been limited to half of the pin 
% radius.
% At higher force values the vast majority of accepted moves are in the 
% direction of the drive, but the largest moves are limited in distance leading
% to the observed saturation.
% The noticeable crossing of the I-V curves at large currents is most likely
% an artifact as well, namely due to the fact that the system is updated 
% locally, rather than globally, in the Metropolis algorithm.  
% At higher vortex densities larger local moves are suppressed by the repulsive
% potential of their nearest neighbors again leading in saturation of the I-V 
% curve.  
% As the density is increased large local moves are further suppressed 
% resulting in I-V curves for systems with a higher vortex density saturating 
% sooner which yields the observed crossing of the I-V characteristics.  

In the following, we mostly report data that have been collected at a driving 
force $f=0.04$.  
This high value was chosen so that washboard noise could be observed over a 
range of accessible vortex densities; however, in some instances the I-V curve 
has already begun to saturate.  
Since we are studying the velocity noise spectrum relative to the mean 
velocity, we do not expect specific artifacts caused by the saturation in the 
voltage noise, nor do we observe any as compared to results obtained at lower 
drive values $f=0.01$ (see the discussion at the end of Sec.~\ref{sec:Columnar 
Defects}).  
%Yet additional investigations may be required to clarify this issue.  

\subsection{Narrowband Noise Characteristics}

\begin{figure}
\begin{center}
% \subfigure[Random Point Defects]{\label{washboardpts}
% \includegraphics[scale=.25]{./washboardpts.eps}}
  \subfigure{\label{washboardpts}
  \includegraphics[scale=.55]{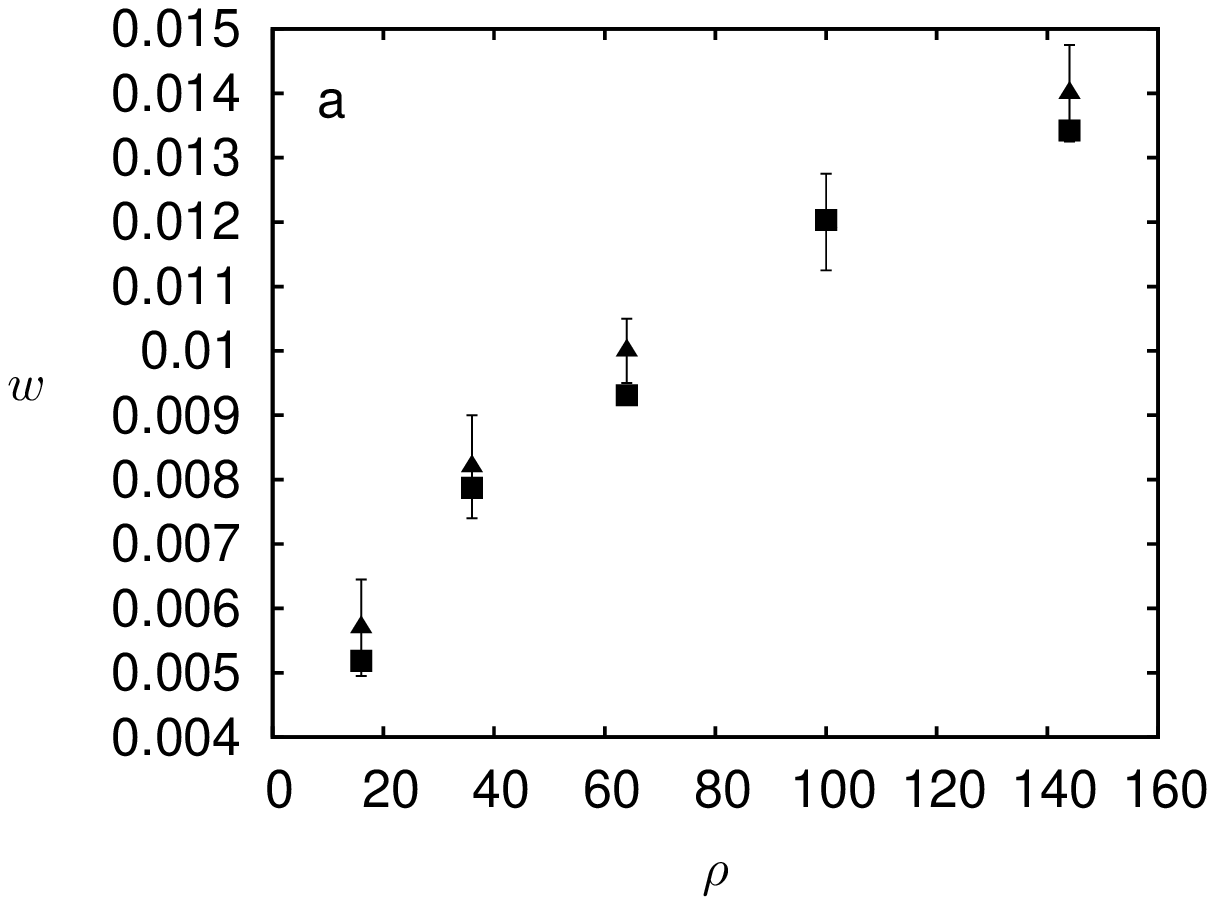}}
% \subfigure[Random Columnar Defects]{\label{washboardran}
% \includegraphics[scale=.25]{./washboardran.eps}}
  \subfigure{\label{washboardran}
  \includegraphics[scale=.55]{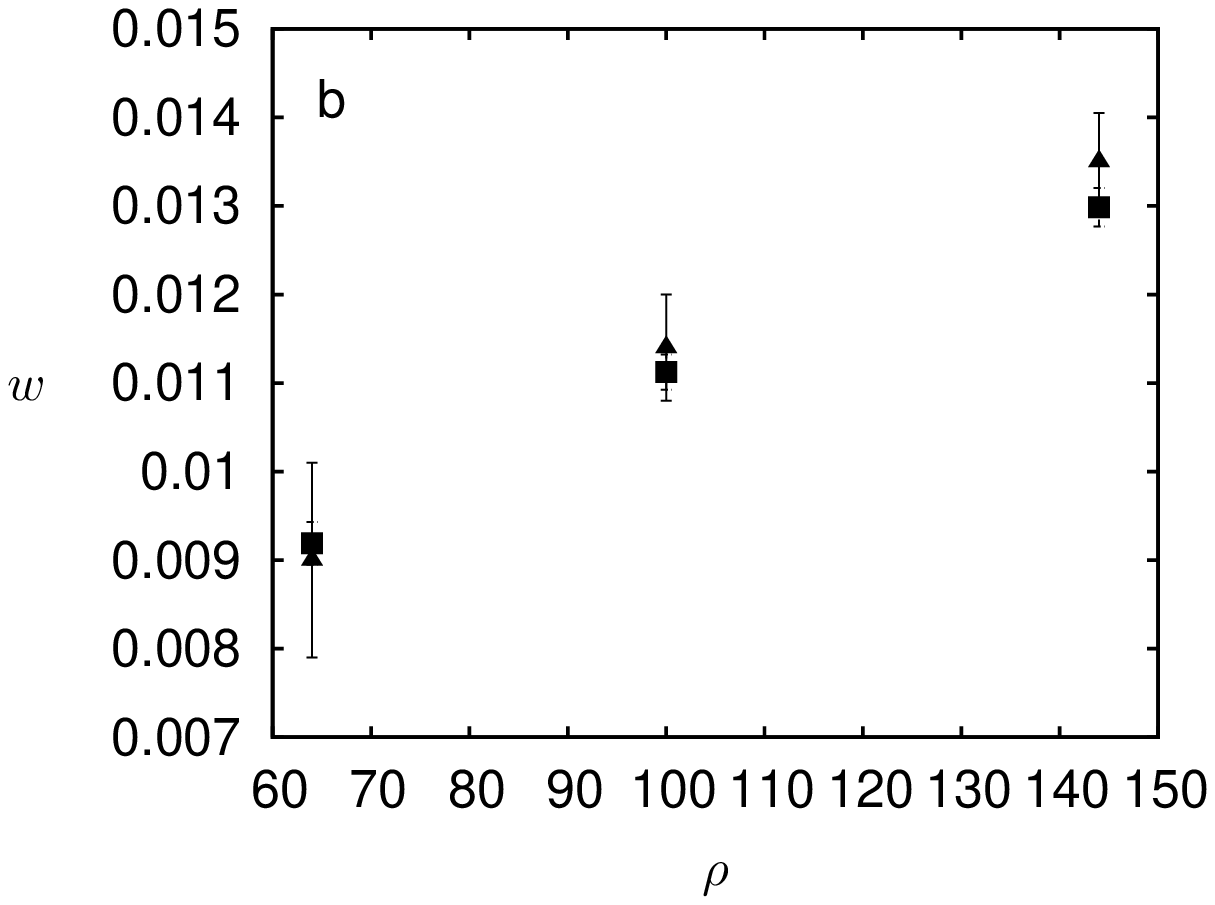}}
\end{center}
\vskip -0.1 truecm
\caption{Washboard frequency versus vortex density $\rho$ for (a) point and (b)
   columnar pinning centers.  
   Vortex density is reported as a count of the number of flux lines in a fixed
   area of size $A = 150 \frac{2}{\sqrt{3}} \, b_0 \times 150 \, b_0$.  
   The results show good agreement between the measured and calculated values 
   for both defect types.  
   The frequency $\omega$ is given in units of rad / MCS.  
   Error bars for the measured values are obtained from the full width at 
   half-maximum of the washboard peaks in Figs.~\ref{random} and \ref{points}. 
   Error bars for the calculated values are smaller than the data points. 
   $\blacktriangle$ - measured, $\blacksquare$ - calculated.}
\label{washboard}
\end{figure}
Power spectral density plots and intensity plots of the structure factor 
$S({\bf k})$ have been obtained for several vortex densities as shown in 
Figs.~\ref{random} and \ref{points}.  
Unless otherwise indicated, all power spectral density plots are 
double-logarithmic with the frequency on the $x$ axis ranging from $0.001$ to 
$0.1$ rad / MCS and power in normalized units on the the $y$ axis in the range
$2.5 \times 10^{-10} \ldots 1 \times 10^{-4}$.
Distinct peaks are observed as well as higher harmonics in many of the power 
spectra for columnar defects.  
Peaks are observed in all of the power spectral plots for point defects.  
In both cases harmonics are always located at integer multiples of the 
fundamental frequency.  
In Fig.~\ref{washboard} the fundamental frequencies are plotted versus the 
number of vortices per unit system size along with the predicted washboard 
frequencies for both point and columnar disorder. 
The washboard frequency is calculated by simply dividing the measured average 
vortex velocity ${\langle v_x \rangle}$ (parallel to the applied drive, and 
averaged over time and defect realizations) by the vortex triangular lattice 
constant.  
Due to the aspect ratio of the system this distance is obtained by dividing the
system length in the direction of the drive ($L_x$) by the square root of the 
number of vortices in the system; hence,
\begin{equation}
\omega = 2\pi {\langle v_x \rangle} \sqrt{N_v} / L_x .
\end{equation}
We obtain good agreement between the measured and predicted values indicating 
that the fundamental frequency in the power spectra plots is indeed the 
washboard frequency.  
Error bars for the measured frequencies are estimated by taking the full widths
at half maximum of the fundamental peaks, while the uncertainties for the 
calculated values are obtained from the standard deviations of the average 
velocities.

\subsubsection{Randomly Distributed Columnar Defects}
\label{sec:Columnar Defects}

\begin{figure}
\begin{center}
% \subfigure[16 lines]{\label{16random}
% \includegraphics[scale=.25]{./16random_xnoise.eps}}
 \subfigure{\label{16random}\includegraphics[scale=.25]{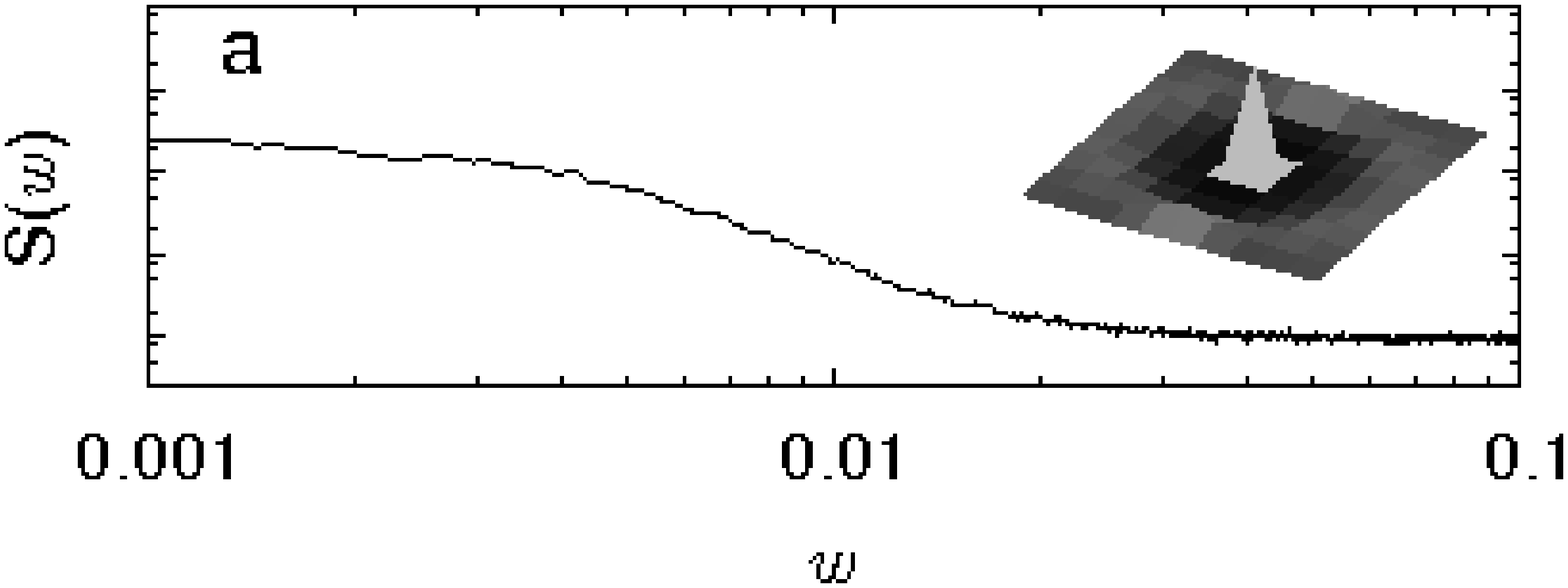}}
 \vskip -0.2 truecm
% \subfigure[36 lines]{\label{36random}
% \includegraphics[scale=.25]{./36random_xnoise.eps}}
 \subfigure{\label{36random}\includegraphics[scale=.25]{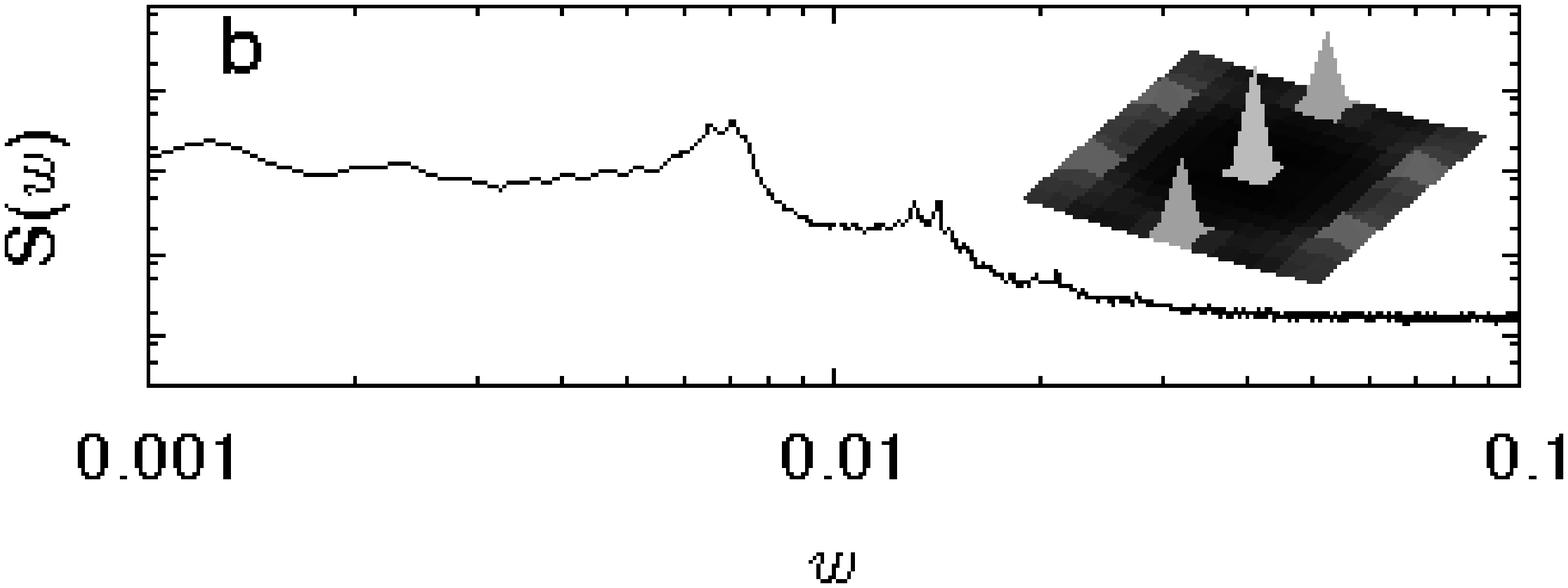}}
 \vskip -0.2 truecm
% \subfigure[64 lines]{\label{64random}
% \includegraphics[scale=.25]{./64random_xnoise.eps}}
 \subfigure{\label{64random}\includegraphics[scale=.25]{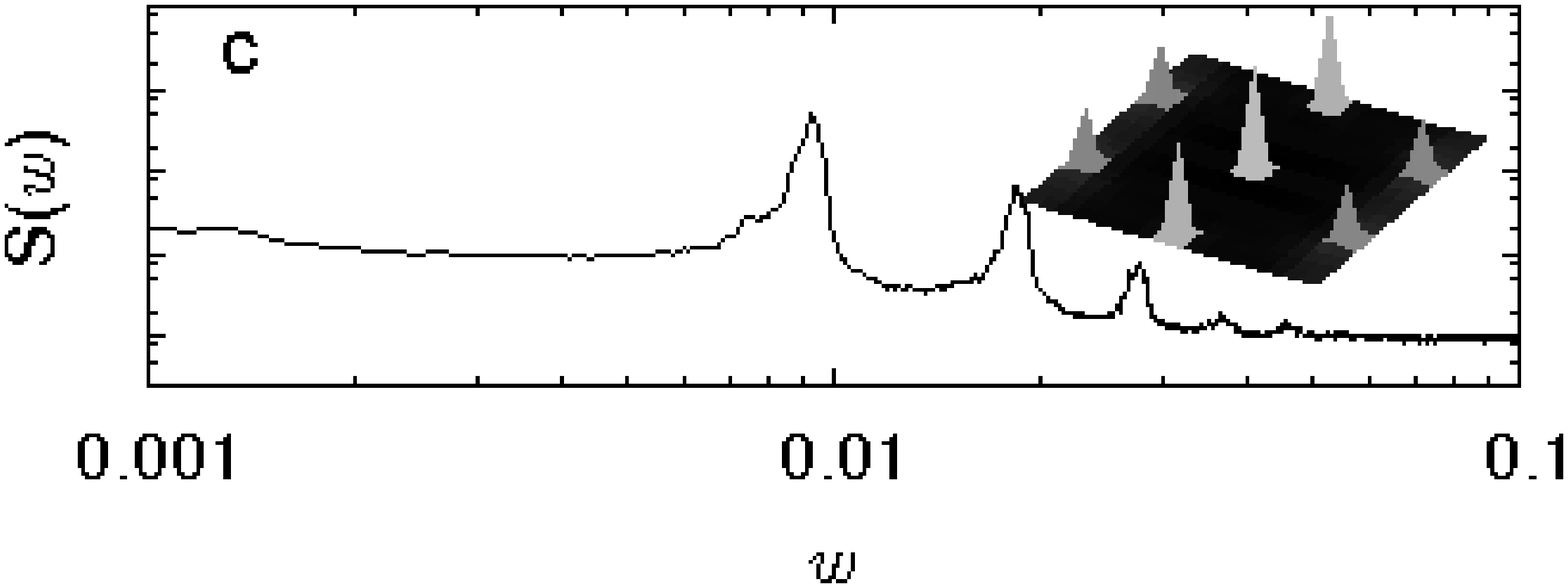}}
 \vskip -0.2 truecm
% \subfigure[100 lines]{\label{100random}
% \includegraphics[scale=.25]{./100random_xnoise.eps}}
  \subfigure{\label{100random}\includegraphics[scale=.25]
  {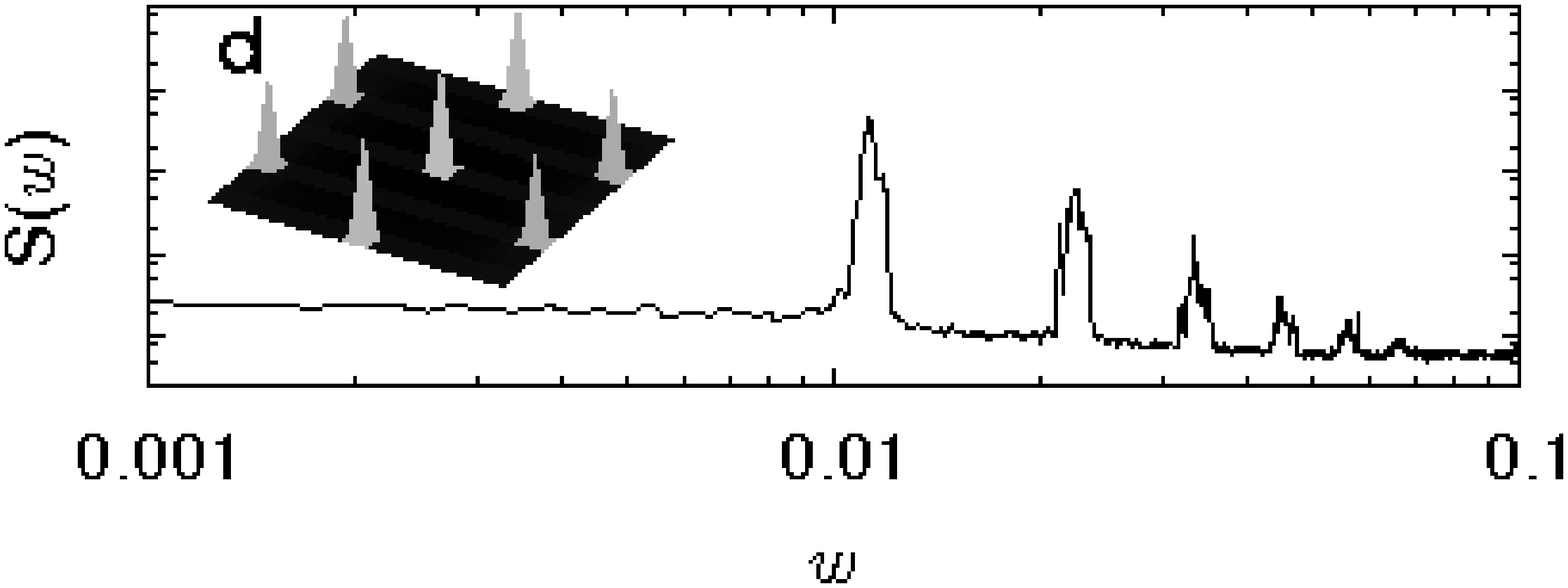}}
% \subfigure[144 lines]{\label{144random}
% \includegraphics[scale=.25]{./144random_xnoise.eps}}
\end{center}
\vskip -0.3 truecm
\caption{Velocity / voltage power spectra measured in the direction of the 
  drive ($x$ direction) and structure factor plots for increasing vortex 
  density in the presence of randomly distributed columnar defects.  
  (a) For a density $\rho = 16 / A = 0.00062/b_0^2$ an isotropic liquid is 
  observed along with a broadband noise signal in the power spectrum.  
  (b) At $\rho = 36 / A = 0.00139/b_0^2$ two (off-center) peaks appear in the 
  structure factor located transverse to the drive direction, accompanied by a 
  narrowband signal in the velocity power spectrum.  
  (c) Six peaks appear in the structure factor plot for 
  $\rho = 64 / A = 0.00246/b_0^2$.  
  The intensities of the peaks with a wave vector component in the $x$
  direction are lower than those with only a $y$ component.  
  In the velocity power spectrum the washboard frequency peak narrows 
  indicating greater temporal coherence, and higher harmonics become more 
  visible.  
  (d) As the density is increased to $\rho = 100 / A = 0.00385/b_0^2$ the 
  structure becomes more fully ordered into a triangular array, and the 
  washboard peak sharpens.}
\label{random}
\end{figure}
\begin{figure}
\begin{center}
  \subfigure[$\rho = 16 / A$]{\label{delauny16random}
  \includegraphics[scale=.35]{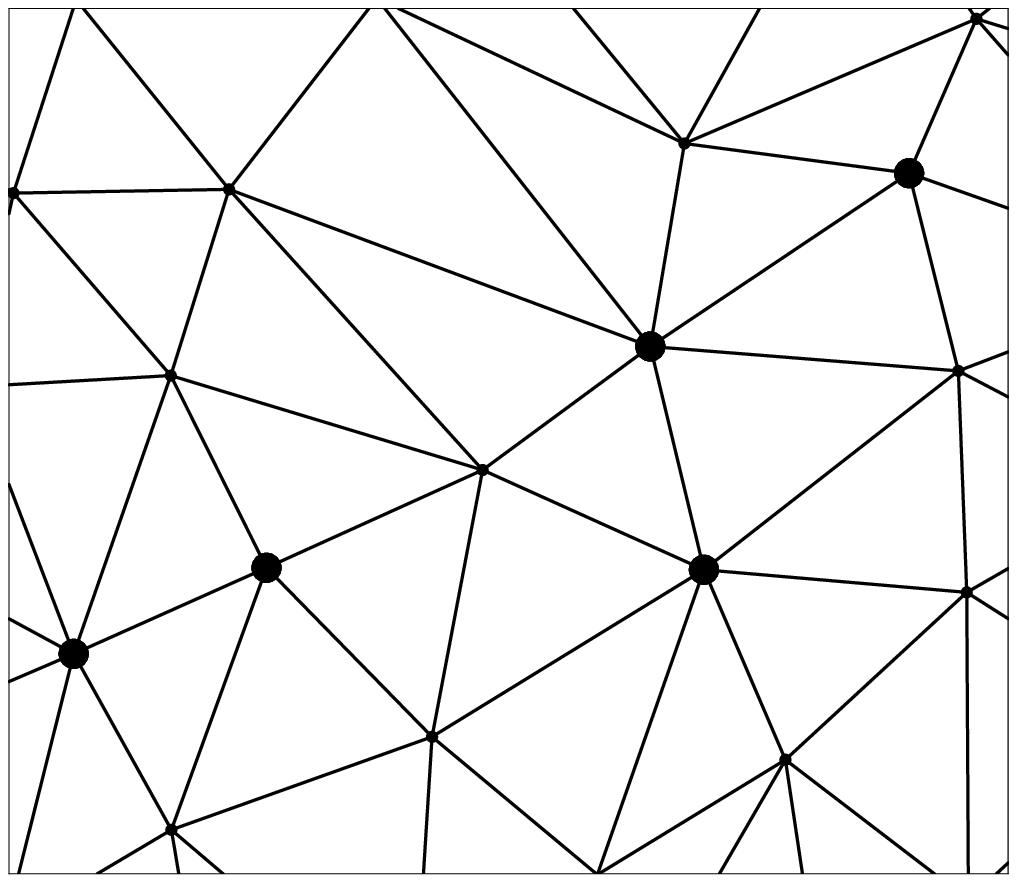}} \
  \subfigure[$\rho = 64 / A$]{\label{delauny64random}
  \includegraphics[scale=.35]{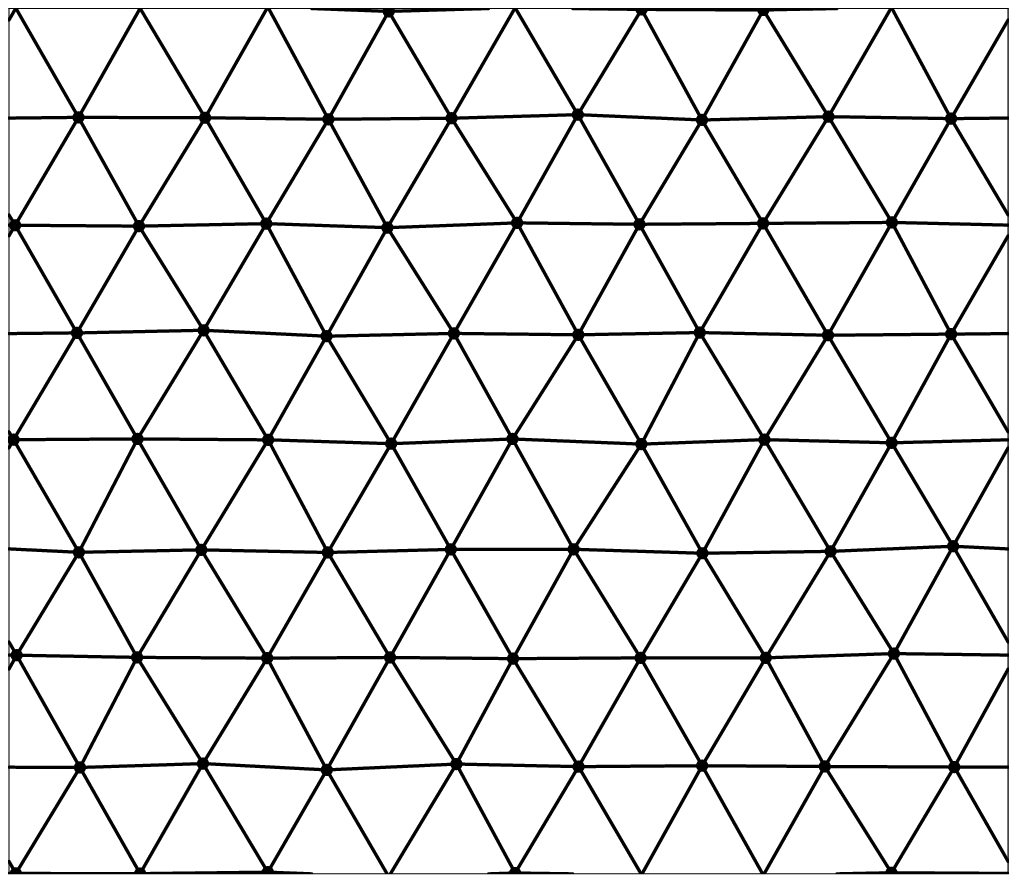}}
  \subfigure[$\rho = 16 / A$]{\label{16ran_traject}
  \includegraphics[scale=.125]{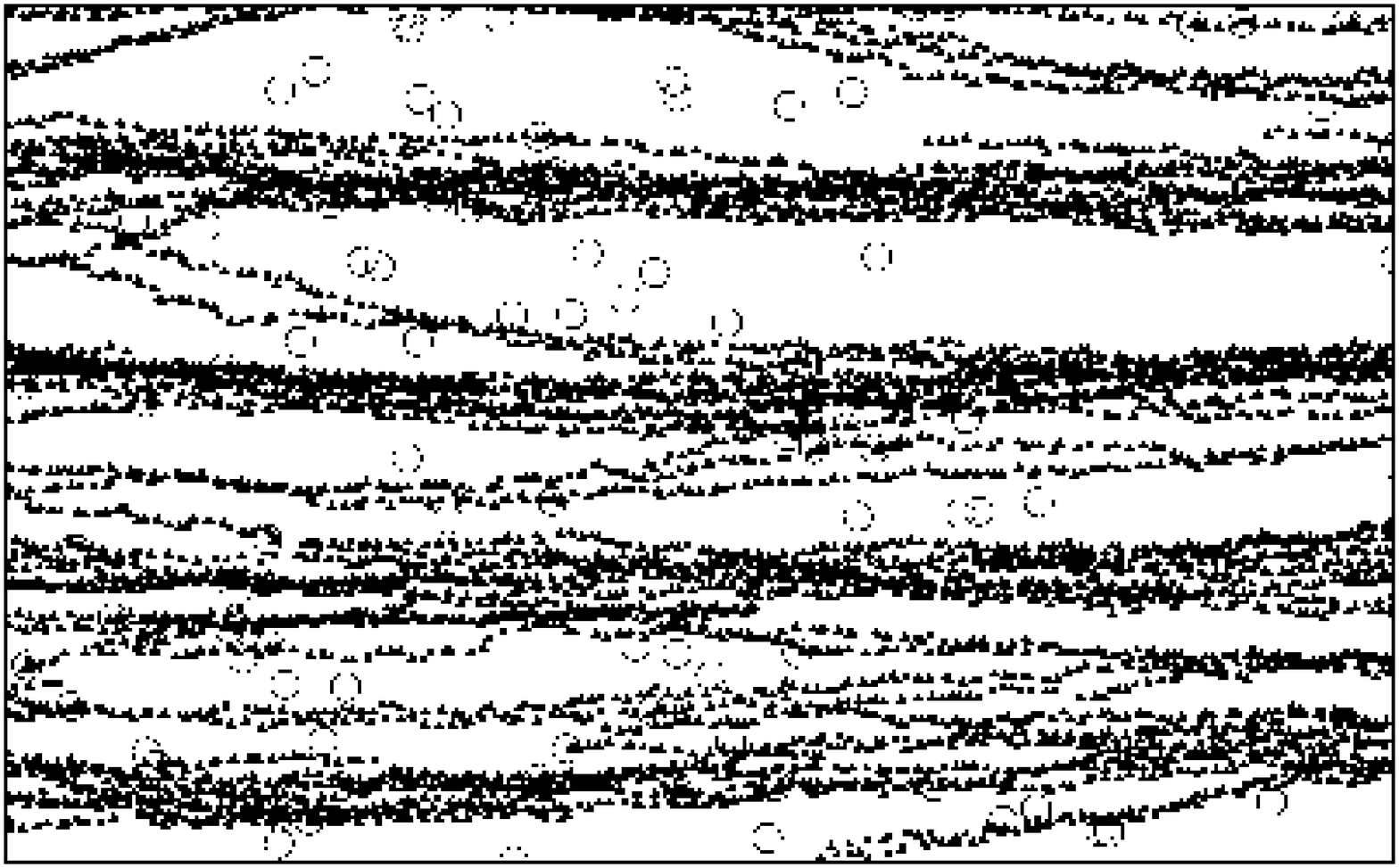}} \
  \subfigure[$\rho = 64 / A$]{\label{64ran_traject}
  \includegraphics[scale=.125]{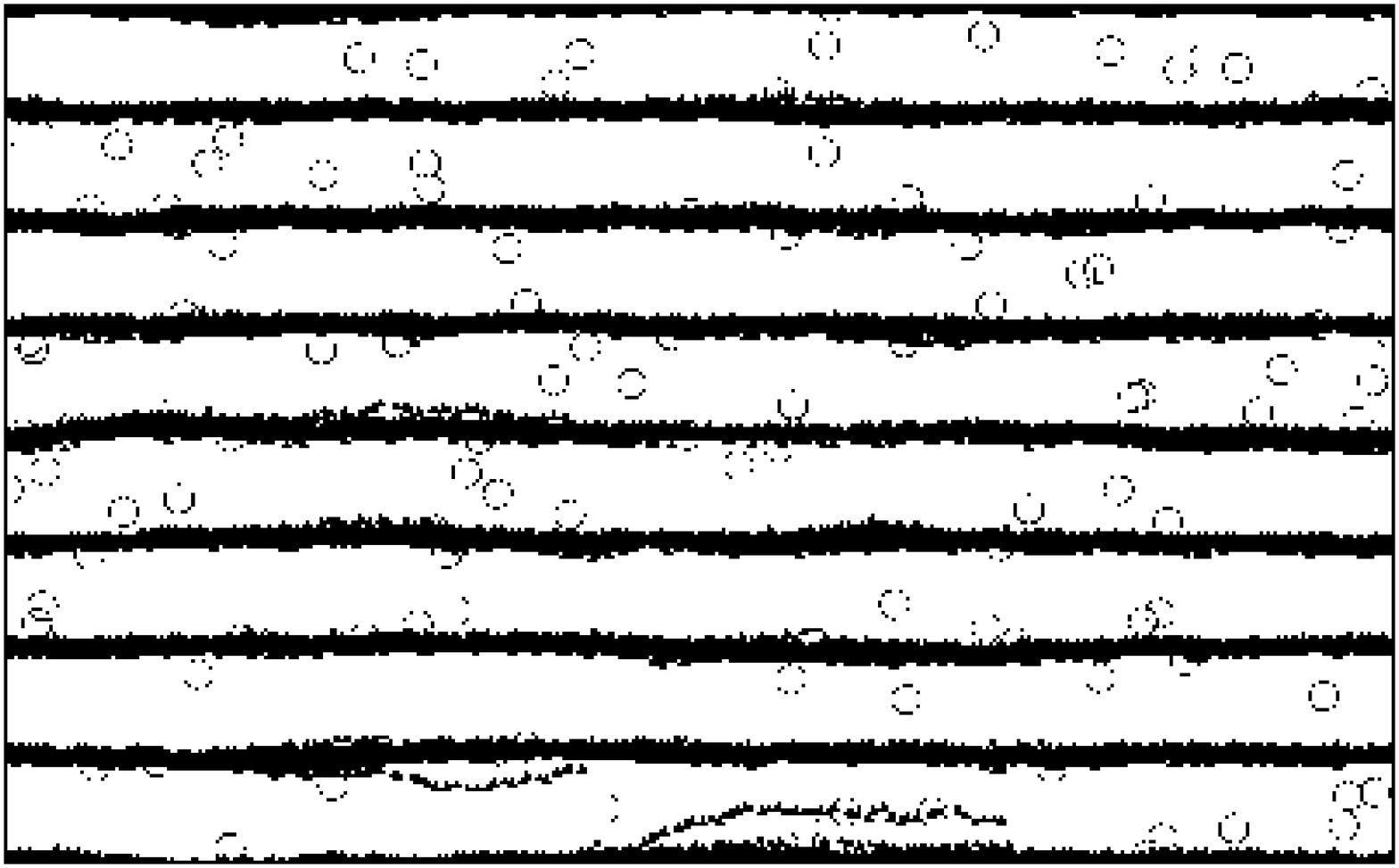}}
\end{center}
\vskip -0.3 truecm
\caption{Delaunay triangulation plots for the positions of (a) $16$ and (b) 
  $64$ flux lines per area $A$ in the presence of random columnar defects.  
  Topological defects are marked.  
  A disordered structure is obtained for flux density $\rho = 16 / A$, while a 
  triangular array free of topological defects is found for $\rho = 64 / A$.  
  A `time exposure' for densities $16 / A$ and $64 / A$ is plotted in (c) and 
  (d) to illustrate the flux line motion.  
  Open circles indicate columnar defect locations, while black lines trace the 
  trajectories of the average position of each vortex.  
  For $\rho = 16 / A$, the vortex trajectories cross suggestive of plastic 
  flow, while for $\rho = 64 / A$ parallel channels form.}
\label{delaunyrandom}
\end{figure}

{\em Noise characteristics and structure.}
Results for flux lines of different densities interacting with random columnar 
defects are displayed in Fig.~\ref{random}.  
With the average spacing between defects set to $15 \, b_0$ the number of 
columns in a unit area $A = 150 \frac{2}{\sqrt{3}} \, b_0 \times 150 \, b_0$ is
$115$.  
For a flux density $\rho = 16 / A$ (a filling fraction of $\sim 1/7$) only 
broadband noise is observed in the velocity power spectrum, see 
Fig.~\ref{16random}.  
The corresponding diffraction pattern shows a ring indicating an isotropic 
liquid or amorphous solid.  
A typical Delaunay triangulation for the vortex positions in a snapshot of a 
particular run with $16$ lines is depicted in Fig.~\ref{delauny16random}.  
The plot shows the presence of a number of topological defects in the vortex 
system.  
The `time exposure' plot in Fig.~\ref{16ran_traject} of the flux line 
center-of-mass positions produces trajectories reminiscent of the `braided 
rivers' observed in two-dimensional plastic flow simulations \cite{OLSON2}.  
The trajectories appear to move and cross within winding channels.  
For a density $\rho = 36 / A$ the two peaks located perpendicular to the drive 
direction fully emerge, suggestive of the predicted moving transverse glass, 
see Fig.~\ref{36random}.
Small intensity peaks with wave vector components parallel to the drive are 
also visible.

As the vortex density $\rho$ is increased to $64 / A$, peaks with an $x$ 
component in the diffraction plot emerge more prominently and sharpen as shown 
in Fig.~\ref{64random}. 
Just as for $\rho = 36 / A$, these peaks are smaller than those located
perpendicular to the direction of the drive.  
In a typical associated Delaunay plot, Fig.~\ref{delauny64random}, topological 
defects in the vortex lattice disappear, and the flux line trajectories form 
parallel channels, running in comparatively straight lines, 
Fig.~\ref{64ran_traject}.  
At $\rho = 100 / A$ the structure plot reveals a well-ordered lattice of flux
lines, see Fig.~\ref{100random}.

These results are a good demonstration of the competing energy scales in the 
system.  
The spatially randomly distributed pinning sites favor disordered flux line 
structures while the vortex repulsion induces ordering into a regular array.  
For $\rho = 16 / A$ the structure factor displays a random vortex 
configuration, indicating that the lattice structure is dominated by the
columnar pinning sites; indeed, the system is close to the depinning threshold,
see Fig.~\ref{depinthr}. 
Individual vortices are pinned for periods of time that are long compared to 
the time it would take for the flux lattice to move one lattice constant, thus 
preventing the formation of the regular structure.  
At density $36 / A$, the system has moved away from the depinning threshold, 
and the repulsive forces begin to separate the vortices into parallel channels 
resulting in spatial periodicity in the $y$ direction (perpendicular to the 
drive), and by reaching $\rho = 64 / A$ the system of moving vortices is 
dominated by the repulsive vortex interaction potential. 
At this stage individual channels begin to couple as additional periodicity 
appears in the $x$ (drive) direction.  
Distinct peaks emerge in the diffraction plot with the peaks possessing a wave 
vector component in the $x$ direction smaller than those with only a $y$ 
component due to weaker coupling.  
As the density is increased to $\rho = 100 / A$ the repulsive energy between 
the vortices grows further leading to a stronger coupling between transverse 
channels and eventually a symmetric triangular lattice structure emerges.

\begin{figure}
\begin{center}
  \includegraphics[scale=.55]{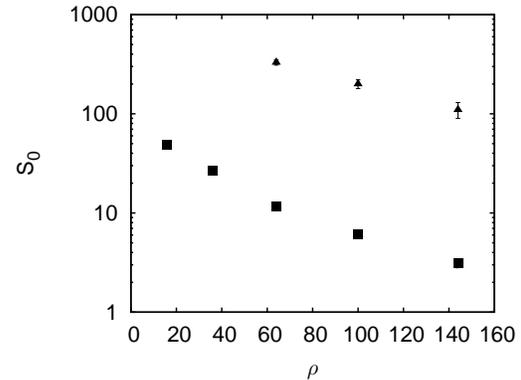}
\end{center}
\caption{Washboard peak power for increasing vortex density for columnar 
  (square data points) and point defects (triangles).  
  In either case the peak intensity is observed to decrease as the flux density
  is increased.}
\label{powertrend}
\end{figure}
The spatial structure of the vortex array is reflected in the associated 
velocity or voltage power spectrum.  
At $\rho = 16 / A$ a broadband signal is visible corresponding to the isotropic
vortex distribution and their consequent random incoherent motion shown in 
Fig.~\ref{16ran_traject}.
Similar broadband noise that is associated with incoherent flux transport has 
been experimentally and numerically observed in a number of studies 
\cite{OLSON2,VEST,TOGAWA}. 
For $\rho = 36 / A$, as the vortices begin to arrange into a lattice a 
narrowband peak appears corresponding to the $x$ wave vector components that 
appear in the structure factor plot.  
At a density $64 / A$ the washboard signal narrows as the vortex system becomes
more structured.  
A sharper peak indicates a greater temporal coherence between the vortices 
which is not surprising; as the density of the vortices is increased each 
vortex is more tightly held in its place and each line `stiffened' by the
repulsive forces exerted by its neighbors.  
For similar reasons the remnants of the broadband noise flatten out more.
This trend continues as the flux density is further increased to $100 / A$.  
As the density is increased from $\rho = 64 / A$ upward we observe a decrease 
in the washboard peak power, as shown in Fig.~\ref{powertrend}.  
The decrease of the fundamental peak intensity is due to the increase in 
stiffness of the lattice structure with growing vortex density.  
As each flux line passes through a defect site it will be less affected by the
pin in a denser system resulting in smaller velocity fluctuations and hence 
reduced power output.

%\begin{figure}
%\begin{center}
%  \includegraphics[scale=.6]{./increasing_den_ran_ynoise.eps}
%\end{center}
%\vskip -0.2 truecm
%\caption{Velocity noise power spectra measured in the $y$ direction for 
%  increasing vortex density.  
%  The pinning centers are randomly distributed columns.  
%  The power spectrum develops in a manner qualitatively similar to the results
%  in the $x$ direction.}
%\label{randomy2}
%\end{figure}

\begin{table*}
\caption{Ratio of the three largest peaks observed in the vortex velocity power
  spectrum in the presence of randomly distributed columnar defects.  
  For each vortex density the ratios of the intensities of the second and third
  peak to the first are reported, for measurements taken in both the $x$ and 
  $y$ directions.  
  The number of runs over which the power spectral density plots were averaged 
  is also listed.
  The corresponding power spectra peak ratios for a periodic piecewise linear 
  (sawtooth) signal is included for comparison.}
\begin{center}
\begin{tabular}{|c|c|c|c|}
\hline vortex density $\rho$ & runs & ratio ($x$ direction) & 
ratio ($y$ direction) \\
\hline $64 / A = 0.00246/b_0^2$ & $115$ & $1$ : $0.20\pm0.02$ : $0.030\pm0.003$
& $1$ : $0.25\pm0.03$ : $0.079\pm0.009$ \\
\hline $100 / A = 0.00385/b_0^2$ & $44$ & $1$ : $0.31\pm0.04$ : $0.055\pm0.008$
& $1$ : $0.30\pm0.05$ : $0.070\pm0.010$ \\
\hline $144 / A = 0.00554/b_0^2$ & $44$ & $1$ : $0.38\pm0.08$ : $0.120\pm0.030$
& $1$ : $0.36\pm0.06$ : $0.055\pm0.008$ \\
\hline $f(x)=x, \, 0<x<2\pi$ && $1$ : $0.25$ : $0.11$ & \\ \hline
% \hline$f(x)=x^2, \, -\pi<x<\pi$ && $1$ : $0.0625$ : $0.012$ & \\ \hline
\end{tabular}
\end{center}
\vskip -0.2 truecm
\label{columnpower}
\end{table*}

{\em Washboard peak harmonics.}
In Table~\ref{columnpower} the ratio of the intensities of the first and second
harmonic with respect to the fundamental peaks are recorded for the velocity 
noise measured in the $x$ and $y$ directions.  
These ratios vary as the flux density increases indicating a change in the 
shape of the velocity vs. time trace.  
It is observed that the ratios approach values similar to those measured for 
point defects (listed in Table~\ref{pointpower} below).  
We will investigate the relationship between the underlying defect type and the
harmonics ratios in Sec.~\ref{variablepinstrength}.  
% For comparison we note the ratio of the fundamental to the first two
% harmonics of the power spectrum for a saw-tooth wave ($f(x)=x,0<x<2\pi$), 
% namely $1 : 1/4 : 1/9$. For the periodic piecewise parabolic signal 
% $f(x)=x^2,-\pi<x<\pi$, the corresponding ratios are $1 : 1/16 : 1/81$.
A narrowband signal is also measured in the $y$ direction transverse to the 
drive for densities of $64 / A$ through $144 / A$ flux lines (not shown).  
%Plots of the transverse velocity noise spectra are displayed in 
%Fig.~\ref{randomy2}.  
The ratios of the harmonics to the fundamental peak intensity in the $y$ 
direction are also reported in Table~\ref{columnpower}.  
The frequency of the fundamental peak as well as the higher harmonics are 
identical to the frequencies measured in the $x$ direction.  
While the power is lower in the $y$ direction (by factors of $10$ to $40$), the
ratios of the peak intensities are similar.  
Since there is no drive in the $y$ direction, these observations could perhaps
be interpreted as a suppression of transverse motion occurring at the same 
frequency as the washboard motion.  
One likely explanation is that as a vortex becomes temporarily trapped in a 
pinning potential, fluctuations transverse to the motion are suppressed until 
the vortex departs from the pinning site, resulting in periodic behavior.

\begin{figure}
\begin{center}
  \subfigure{\label{radgran1}\includegraphics[scale=.55]{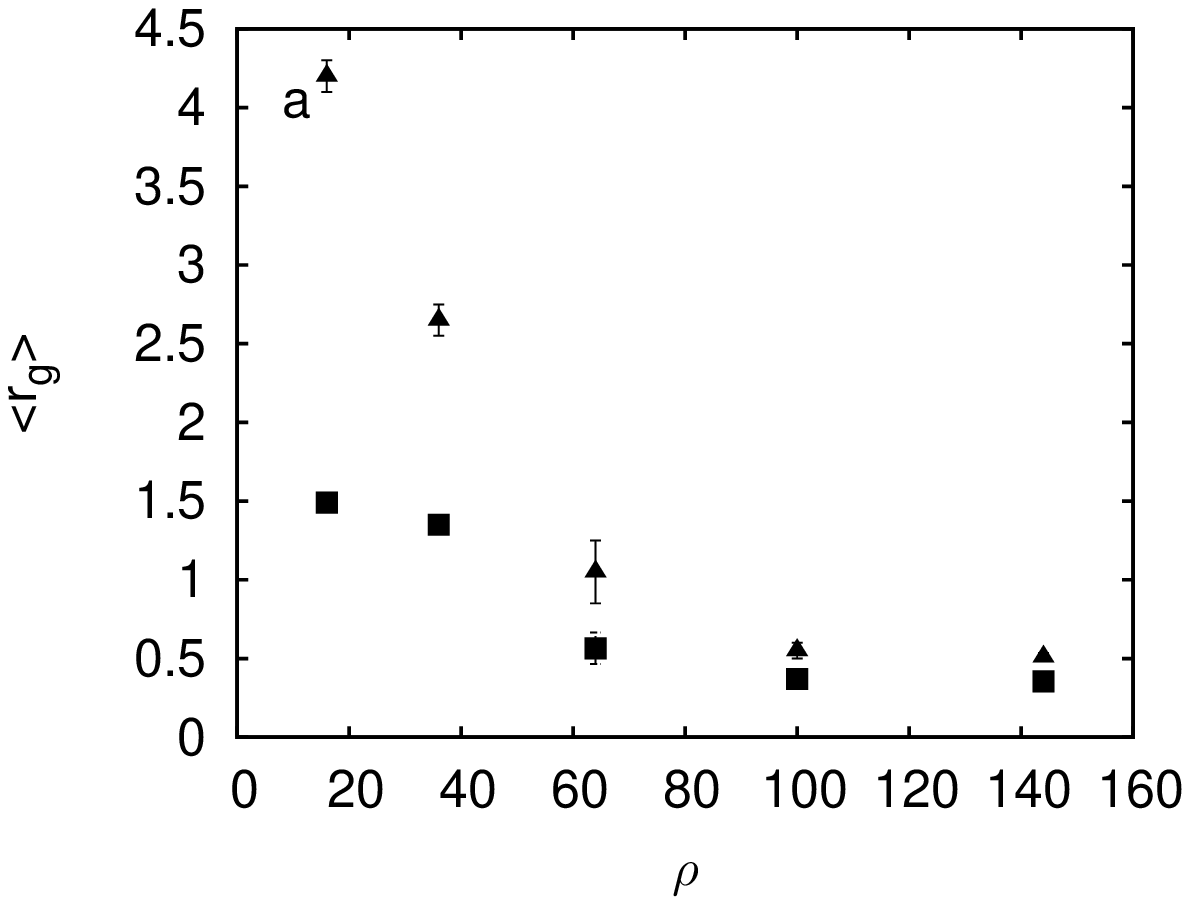}}
  \subfigure{\label{radgpts}\includegraphics[scale=.55]{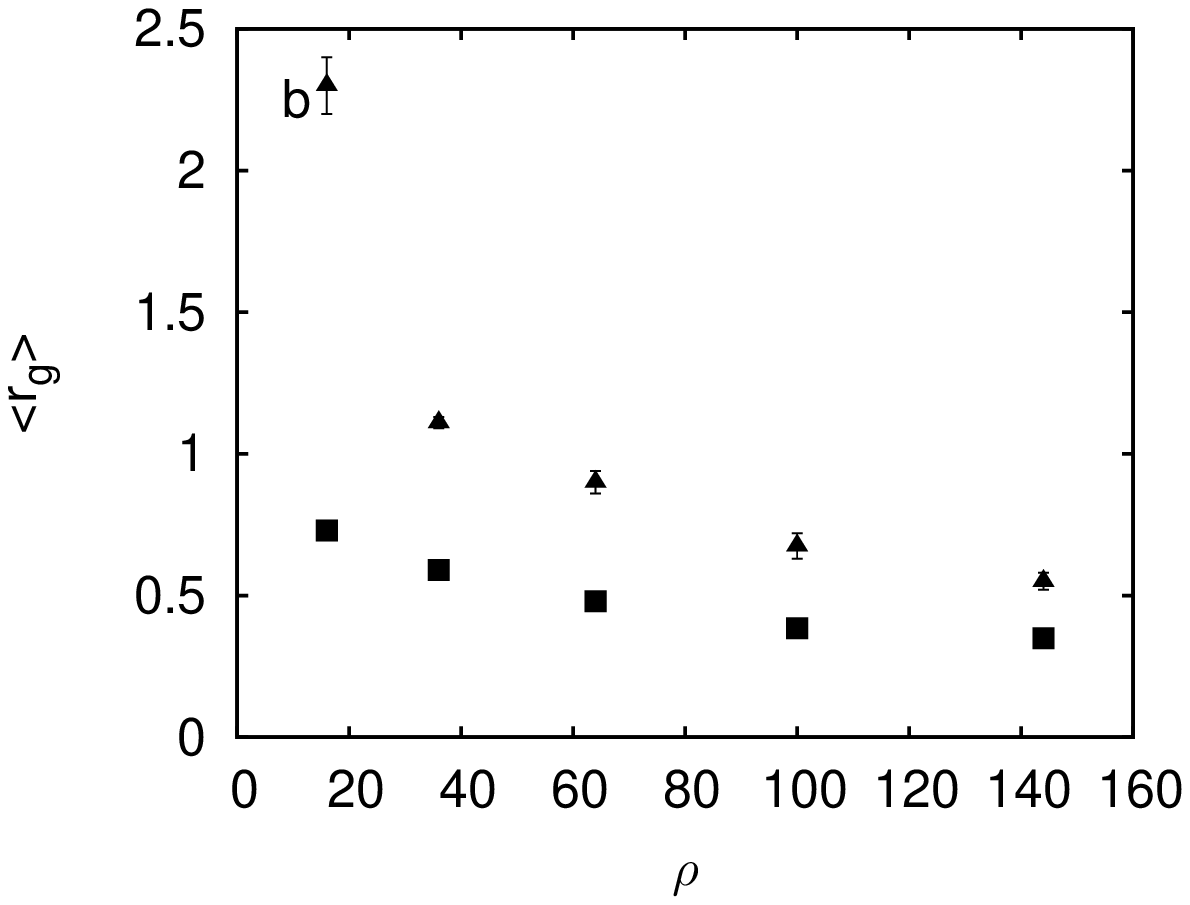}}
\end{center}
\vskip -0.1 truecm
\caption{Components of the mean radius of gyration for (a) columnar and (b) 
  point pins.  
  For both defect types the radius of gyration decreases with increasing vortex
  density.  
  At higher flux densities vortices are caged by nearest neighbors, suppressing
  elastic flux line stretching.  
  Lengths are reported in units of the pin radius $b_0$.  
  $\blacktriangle$ - $x$ component, $\blacksquare$ - $y$ component.}
\label{radgran}
\end{figure}
{\em Radius of gyration.}
In order to obtain additional information about the three-dimensional shape of 
the elastic vortex lines moving through a sample with columnar defects, we have
obtained radius of gyration data, averaged over vortices, time, and defect 
configurations.  
We interpret the radius of gyration as the distance the elastic line is 
stretched from its average (center-of-mass) position.  
The components of the radius of gyration in the $x$ and $y$ direction vs. flux 
density are plotted in Fig.~\ref{radgran1}.  
As the flux density is increased the radius of gyration decreases indicating 
that the vortex lines are straightened at higher densities due to the stronger 
repulsion exerted by their nearest neighbors.  
The data also display anisotropy in the flux line `stretching'.  
The magnitude of the radius of gyration is markedly greater along the direction
of the drive than perpendicular to it, with the $x$ and $y$ components 
approaching each other only at quite large densities.  
For dilute systems we would expect the competition between the drive and the 
pinning potential to result in flexible vortices depinning in sections, with 
some parts of the flux lines leaving the columnar pins via `double-kink' and
`half-loop' saddle-point configurations \cite{NEL1}. 
In the presence of the drive the free flux line sections move forward and grow,
stretching the vortex until it depins completely.  
Since there is no drive in the $y$ direction, the $y$ component of the radius 
of gyration is smaller; however, the same trend is observed with increasing 
density as with the $x$ component.  
At high densities the vortices are stiffer due to a smaller vortex separation 
distance.  
Rather than stretching, depinning tends toward an `all or nothing' process with
the tightly packed vortex ensembles effectively becoming two-dimensional.

\begin{figure}[b]
\begin{center}
  \subfigure{\label{144 low drive}
  \includegraphics[scale=.25]{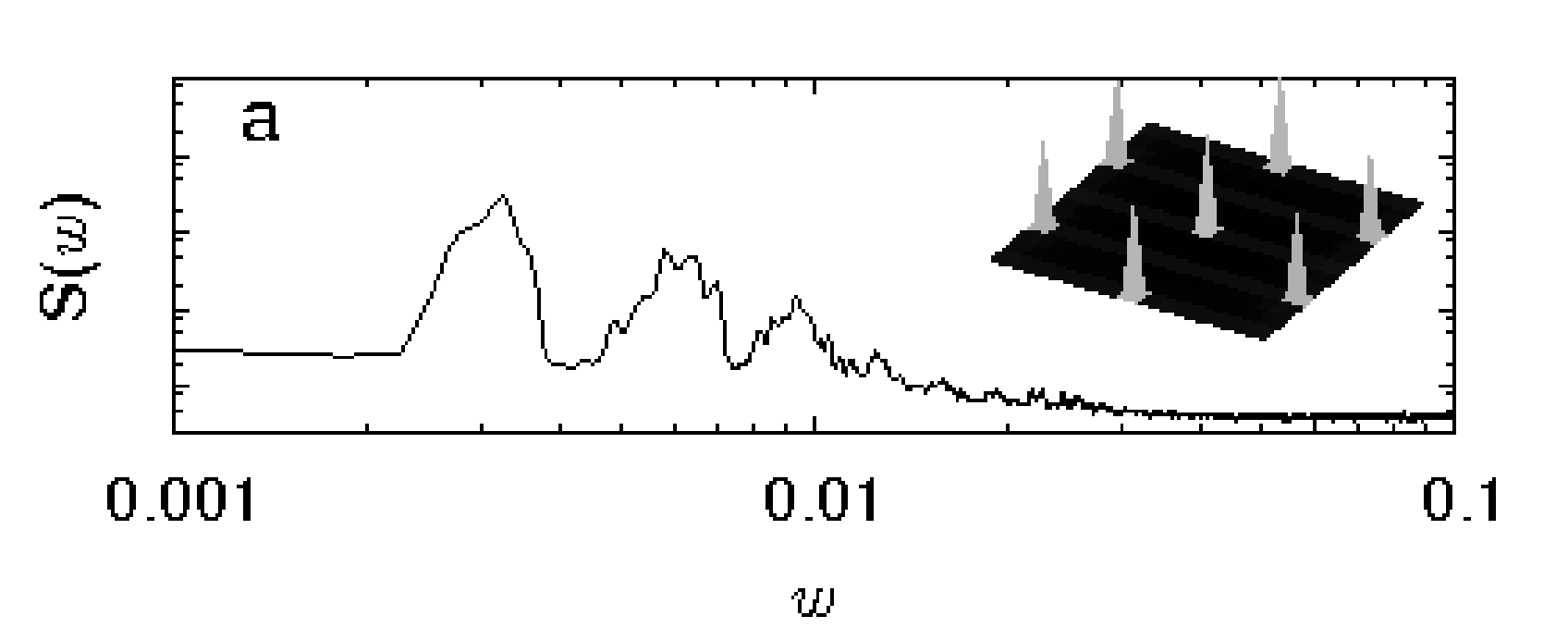}}
  \vskip -0.2 truecm
  \subfigure{\label{144 high drive}
  \includegraphics[scale=.25]{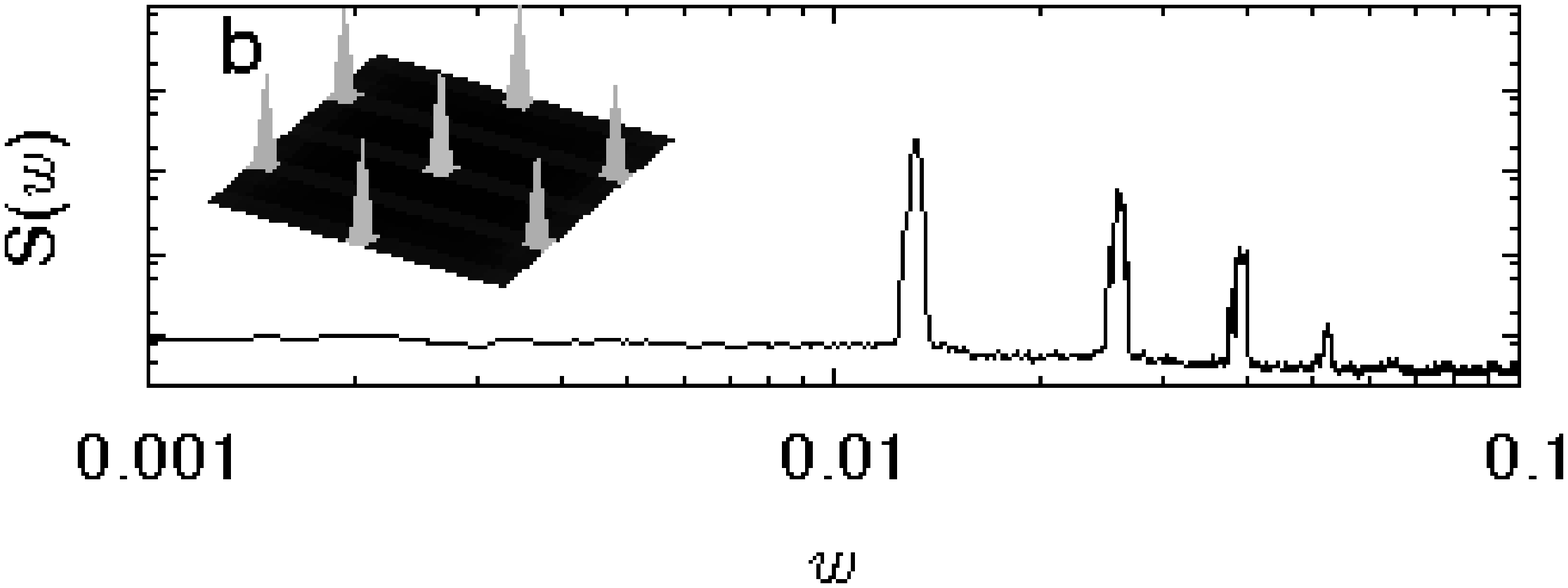}}
\end{center}
\vskip -0.3 truecm
\caption{Voltage power spectra for (a) low ($f=0.01$) and (b) high ($f=0.04$) 
  driving forces at $\rho = 144 / A$ in the presence of columnar defects.}
\label{highlowdrive}
\end{figure}
{\em Comparison with results obtained at small drive.}
The power spectra results reported above have been obtained at a rather high 
applied drive $f=0.04$ chosen so that the washboard peak could be observed over
a range of vortex densities.  
However, as discussed above, the velocity in the I-V characteristics 
(Fig.~\ref{ivresults}) begins to saturate at such high drive values, especially
for systems with higher flux density.   
We believe that this is an artifact of the limited step size in the simulation.
In order to investigate whether the velocity saturation influences the velocity
fluctuations we now examine the power spectra at a lower drive value $f=0.01$
and compare the results to those for $f=0.04$.  
We consider here a dense vortex system with $\rho = 144 / A$ subject to 
randomly placed columnar defects.  
This flux density was chosen to preserve the washboard peak at a lower drive.  
Results are displayed in Fig.~\ref{highlowdrive}.  
These data suggest that the velocity saturation does {\em not} adversely affect
the velocity power spectra.  
The expected drop in washboard frequency corresponding to the decrease in 
applied drive is observed.  
The full peak widths at half maximum of the fundamental are measured at high 
and low drive and the results are found to be comparable 
($0.00036~{\rm rad}/MCS$ and $0.00041~{\rm rad}/MCS$, respectively) indicating 
that the temporal correlations are not affected by the saturation.  
Both spectra display a similar harmonic ratio $1 : 0.40\pm0.05 : 0.10\pm0.01$. 
Likewise the structure factors are found to be similar.  
These results suggest that data obtained at high drive values do not possess 
any pronounced artifacts due to velocity saturation.

\subsubsection{Randomly Placed Point Defects}

\begin{figure}
\begin{center}
 \subfigure{\label{16points}\includegraphics[scale=.25]{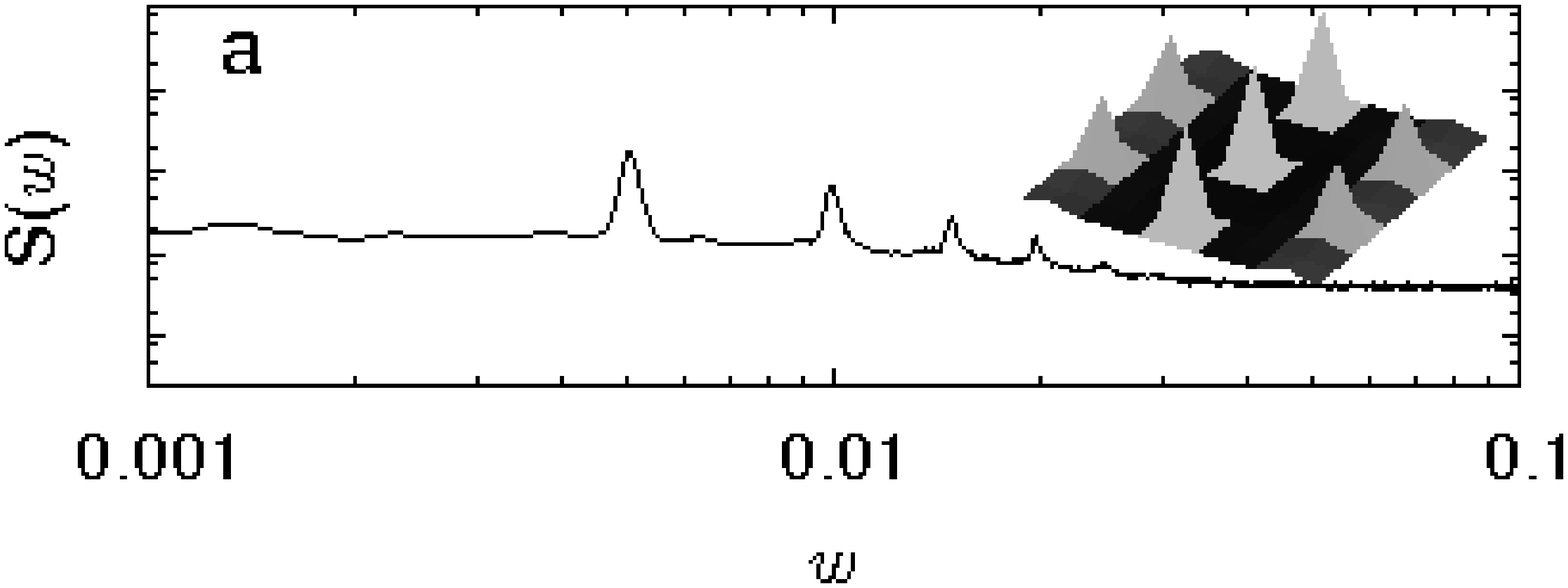}}
 \vskip -0.2 truecm
 \subfigure{\label{36points}\includegraphics[scale=.25]{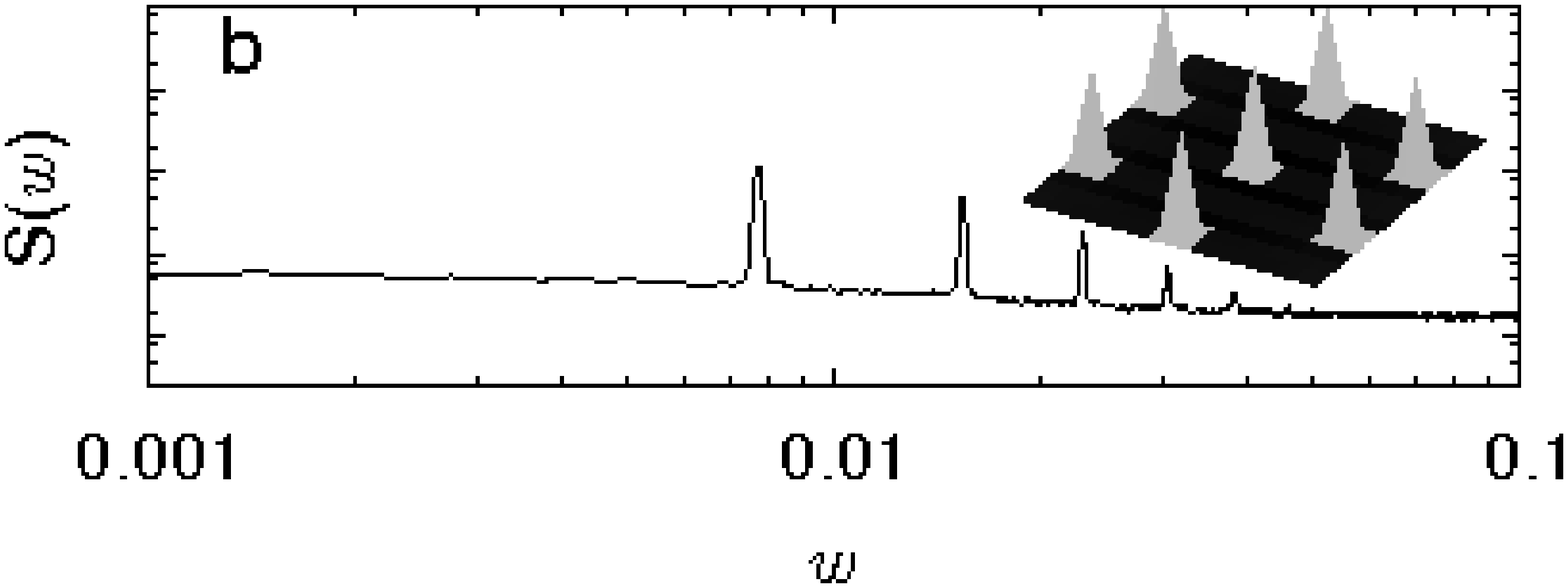}}
 \vskip -0.2 truecm
 \subfigure{\label{64points}\includegraphics[scale=.25]{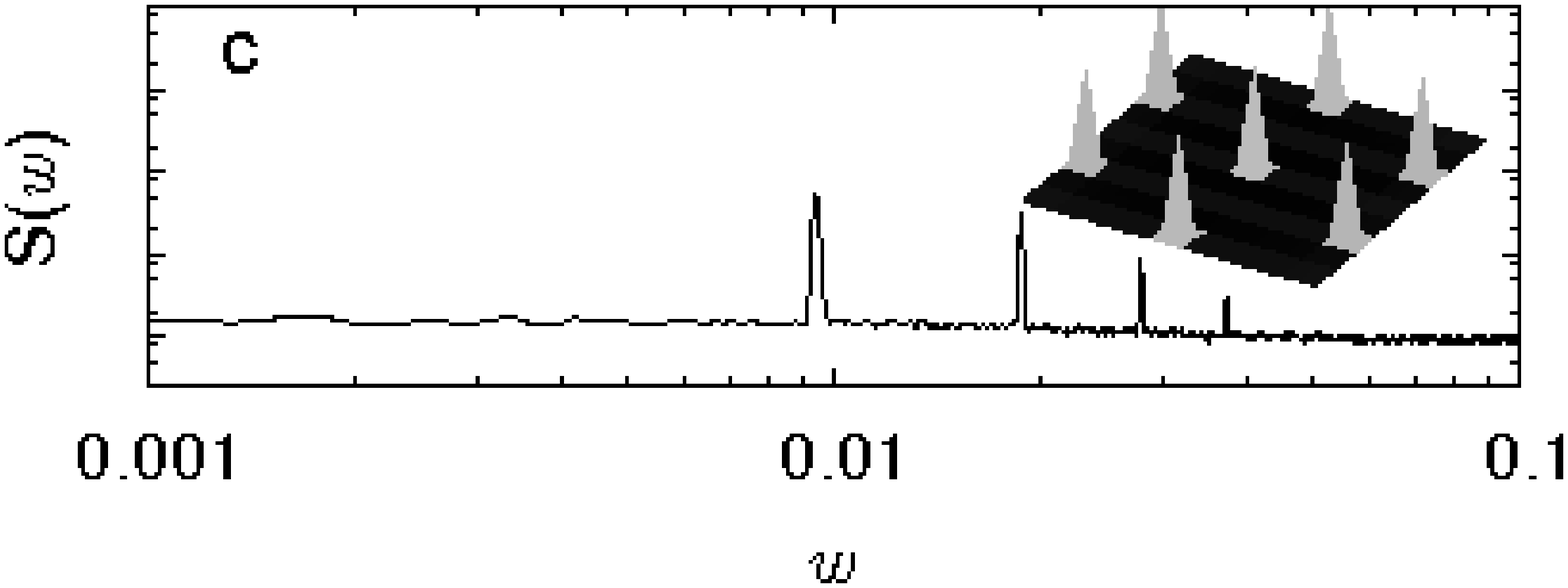}}
 \vskip -0.2 truecm
 \subfigure{\label{100points}\includegraphics[scale=.25]
 {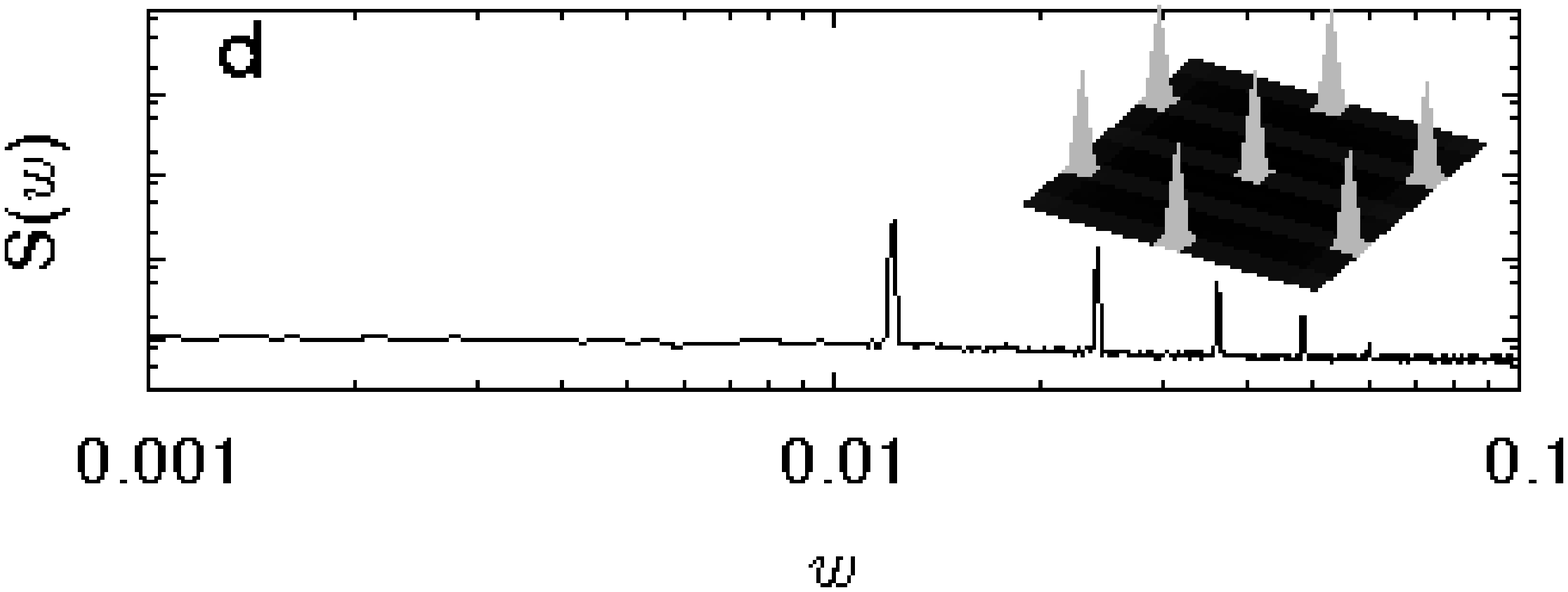}}
% \subfigure{\label{144points}
% \includegraphics[scale=.25]{./144points_xnoise.eps}}
\end{center}
\vskip -0.3 truecm
\caption{Power spectra and structure factor plots for increasing vortex density
  $\rho$ in the presence of point pins, with 
  (a) $\rho = 16 / A = 0.00062/b_0^2$, (b) $36 / A = 0.00139/b_0^2$, 
  (c) $64 / A = 0.00246/b_0^2$, and (d) $100 / A = 0.00385/b_0^2$.  
  Narrowband washboard noise peaks are observed in all thespectral density 
  plots, and six-fold coordination in the corresponding vortex structure
  factors.}
\label{points}
\end{figure}

\begin{figure}[b]
\begin{center}
  \subfigure[$\rho = 16 / A$]{\label{delauny16points}
  \includegraphics[scale=.35]{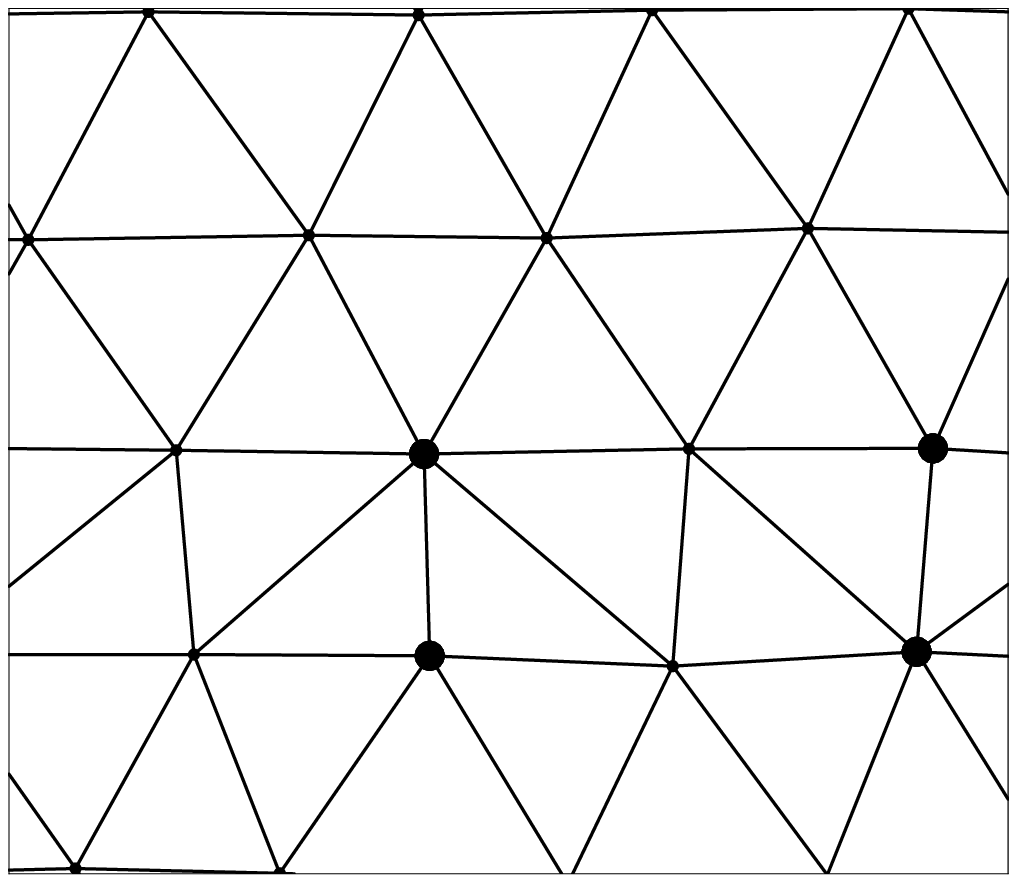}} \
  \subfigure[$\rho = 36 / A$]{\label{delauny36points}
  \includegraphics[scale=.35]{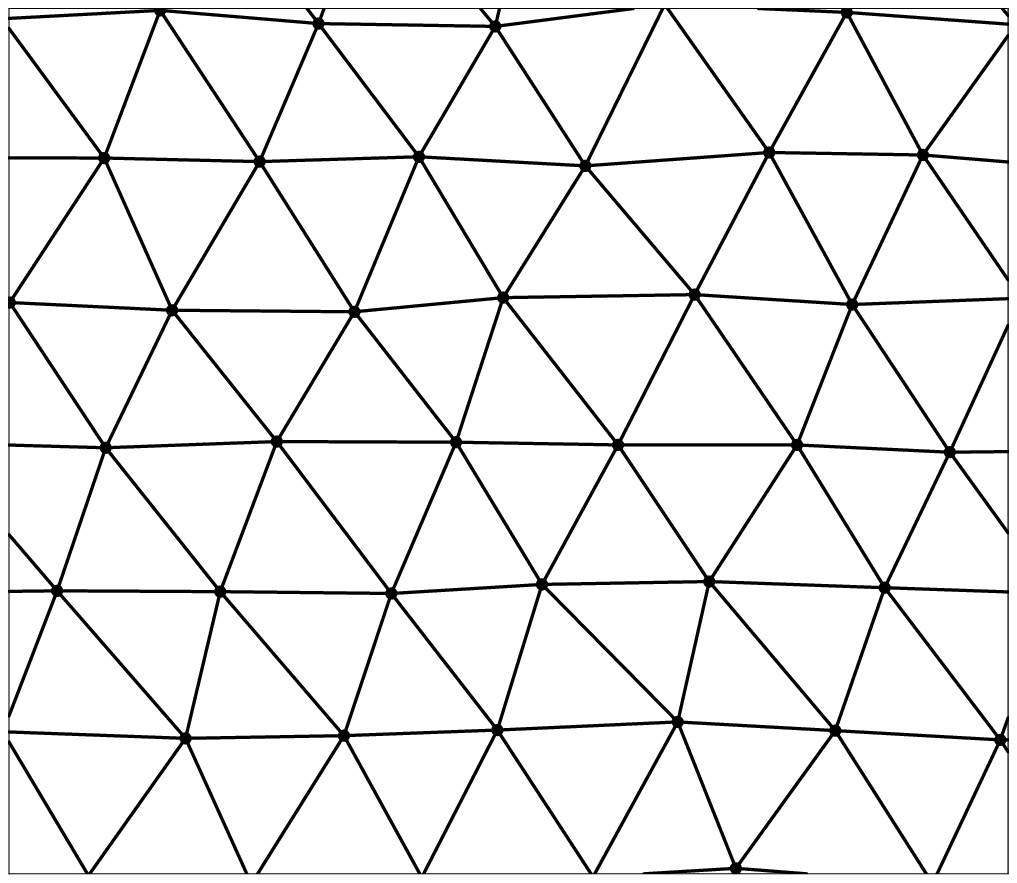}}
  \end{center}
\vskip -0.3 truecm
\caption{Delaunay triangulation for vortices with densities 
  (a) $\rho = 16 / A$ and (b) $36 / A$, interacting with randomly distributed 
  point pinning centers. Topological defects are indicated.}
\label{delaunypoints}
\end{figure}

The results for the velocity power spectra and associated structure factors for
vortices driven through randomly distributed point defects show many 
qua\-litative similarities to those obtained for columnar pins, see 
Fig.~\ref{points}.
However, the structure factor plot associated with an isotropic liquid is never
observed in our simulations, even at the lowest flux densities.  
The intensities of the structure factor peaks increase with growing flux 
density, indicating an increasing degree of positional order in the system.  
For the lowest density system, Fig.~\ref{16points}, the structure factor plot
implies greater spatial periodicity perpendicular to the direction of the drive
than parallel to it, similar to the case of columnar defects, suggesting that 
parallel channels are not well-coupled in this regime.  
This is further evident in the Delaunay triangulation plot shown in 
Fig.~\ref{delauny16points}.  
Compared to the system with a density $16 / A$, the Delaunay plot for 
$\rho = 36 / A$, Fig.~\ref{delauny36points}, yields greater alignment 
perpendicular to the drive owing to the disappearance of topological defects in
the vortex lattice.  
The reason for this is similar to that for columnar defects; the pinning sites 
introduce shear between parallel channels resulting in less-ordered rows of 
vortices that run perpendicular to the drive.  
As the density increases the vortex repulsion becomes the dominant energy in 
the system, and the coupling between parallel channels is enhanced.

In the velocity power spectrum for flux density $\rho = 16 / A$ depicted in 
Fig.~\ref{16points}, the narrowband signal rests on a low broadband background 
signal.  
As the vortex density increases the broadband component diminishes, as do the
power and width of the washboard peak.  
In general, the peak widths are smaller than those observed for columnar 
defects, as to be expected for uncorrelated and less efficient pinning centers.
In the spectra where peaks are resolvable, the fundamental peak and higher 
harmonics are located at frequencies that are identical in the $x$ and $y$ 
directions.
The ratio of the fundamental to the harmonics in either direction are listed in
Table~\ref{pointpower}.  
%A narrowband signal is also measured in the $y$ direction for point defects,
%as shown in Fig.~\ref{pointsy}.  
%In the spectra where peaks are resolvable, the fundamental peak and higher 
%harmonics are located at frequencies identical those measured in the $x$
%direction.
%The power ratios of the harmonics are also recorded in Table~\ref{pointpower}
%for the voltage noise measured in the $y$ direction.  
These ratios for flux densities $\rho = 16 / A$ through $100/A$ lines are 
comparable suggesting a similar shape of the velocity-time traces.  
For a system with $\rho = 144 / A$ the ratio is however markedly different.
At this point it is not clear to us which physical mechanism dictates the 
detailed values of these peak intensity ratios. 
Yet, as we shall further argue in Sec.~\ref{variablepinstrength} below, we 
believe that these ratios characterize the overall deformation of the vortex 
lattice rather than reflect the geometry of the pinning centers.
\begin{table*}
\caption{Ratio of the three largest washboard peaks observed in the velocity 
  power spectrum for point defects.  
  For each vortex density the ratios of the intensities of the second and third
  peak to the first are reported, for measurements taken in both $x$ and $y$
  directions.  
  The number of runs over which the power spectral density plots were averaged 
  is also included.}
\begin{center}
\begin{tabular}{|c|c|c|c|}
\hline vortex density $\rho$ & runs & ratio ($x$ direction) & ratio ($y$ 
direction) \\
\hline $16 / A = 0.00062/b_0^2$ & $44$ & $1$ : $0.67\pm0.09$ : $0.27\pm0.02$ 
& n / a \\
\hline $36 / A = 0.00139/b_0^2$ & $44$ & $1$ : $0.55\pm0.06$ : $0.22\pm0.02$ 
& $1$ : $0.4\pm0.1$ : $0.40\pm0.10$ \\
\hline $64 / A = 0.00246/b_0^2$ & $276$ & $1$ : $0.58\pm0.03$ : $0.24\pm0.01$ 
& $1$ : $0.51\pm0.04$ : $0.16\pm0.01$ \\
\hline $100 / A = 0.00385/b_0^2$ & $88$ & $1$ : $0.56\pm0.05$ : $0.25\pm0.02$ 
& n / a \\
\hline $144 / A = 0.00554/b_0^2$ & $88$ & $1$ : $0.80\pm0.10$ : $0.19\pm0.02$ 
& $1$ : $0.43\pm0.06$ : $0.08\pm0.01$ \\
\hline
\end{tabular}
\end{center}
\vskip -0.2 truecm
\label{pointpower}
\end{table*}

%\begin{figure}[b]
%\begin{center}
%\includegraphics[scale=.6]{./increasing_den_pts_ynoise.eps}
%\end{center}
%\vskip -0.3 truecm 
%\caption{Velocity / voltage noise power spectra measured in the $y$ direction 
%  for increasing vortex density, in a system with randomly distributed point 
%  pins.  
%  The power spectrum develops in a manner qualitatively similar to the results
%  in the $x$ direction.}
%\label{pointsy}
%\end{figure}

The mean flux line radius of gyration (averaged over vortices, time, and 
pinning site distributions) vs. vortex density for point defects is plotted in 
Fig.~\ref{radgpts}.  
The behavior for both components is similar to that of a system subject to 
randomly placed columnar defects.  
A larger radius of gyration is observed for low-density systems, while it 
decreases for both $x$ and $y$ components as the flux density is increased.  
Unlike for columnar defects, the pinning force for uncorrelated point disorder 
does not add coherently over the length of the flux lines.  
Whereas the random spatial distribution of point pins promotes thermal flux
line wandering, the stretching of the vortices while moving through a defect is
not as severe.  
Comparing the data to those for columnar defects with identical flux densities,
we note that the magnitudes of both components of the radius of gyration are 
smaller for point defects. 
As the density is increased both components of the radius of gyration for point
defects approach the same values as also seen in the corresponding curves for 
columnar defects.

\begin{figure}
\begin{center}
  \subfigure{\label{64random_long} 
  \includegraphics[scale=.25]{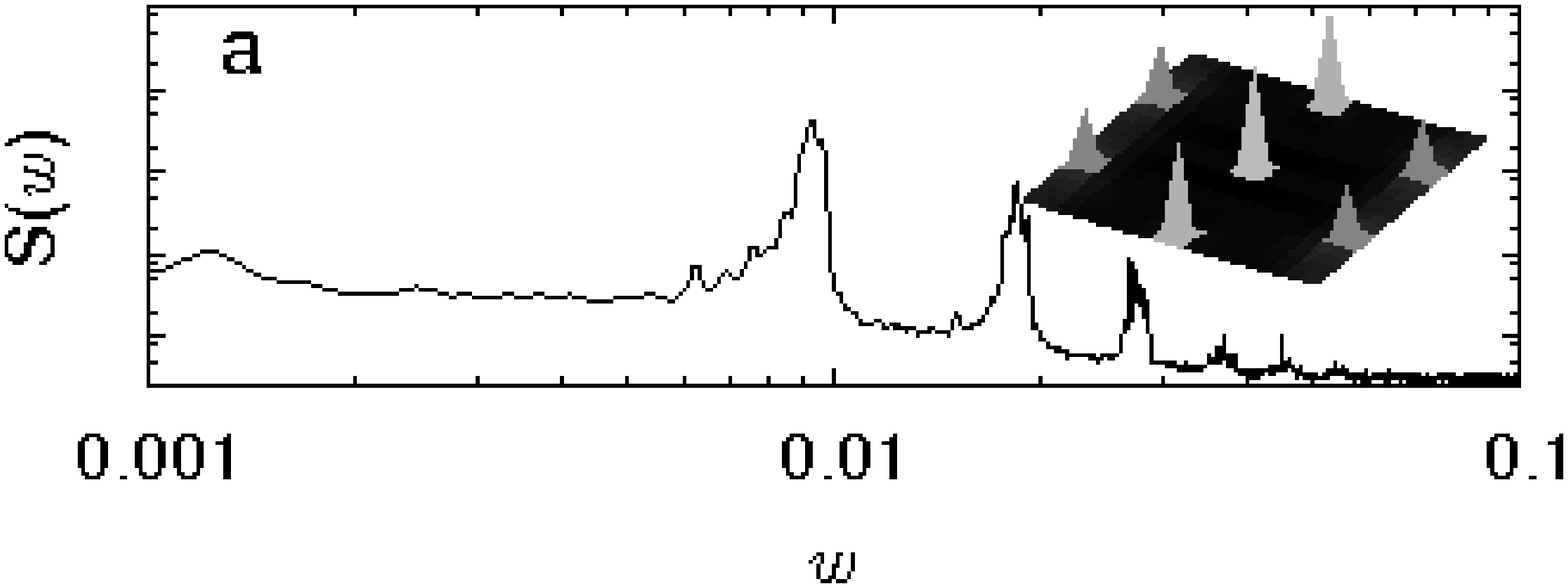}}
  \vskip -0.2 truecm
  \subfigure{\label{64points_long}
  \includegraphics[scale=.25]{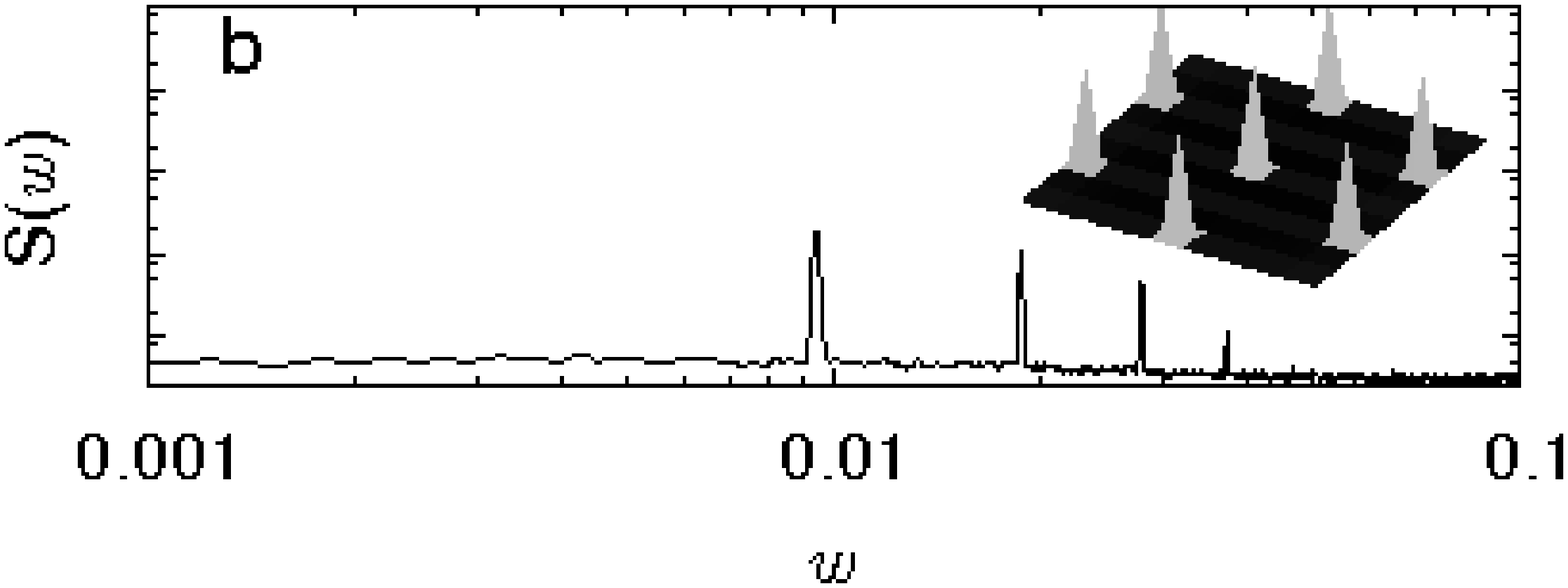}}
\end{center}
\vskip -0.3 truecm
\caption{Velocity fluctuation power spectra and structure factor plots for a 
  flux density $\rho = 64 / A$ for (a) columnar and (b) point defects in a 
  spatially extended system with $L_z = 60$, i.e., all flux lines have been 
  lengthened by a factor of $3$ compared to the previous runs.  
  The results for columnar defects shown in (a) are very similar to those for 
  the same density with shorter lines, as depicted in Fig.~\ref{64random}.  
  On the other hand, for point defects (b) the power drops noticeably compared 
  to Fig.~\ref{64points}.}
\label{64long}
\end{figure}

We have also examined the effect of the total vortex length $L_z$ on the 
velocity power spectrum.  
Data for $L_z = 60$ (three times the length of the vortices shown in all 
previous results) at a density of $64$ lines have been included in 
Figs.~\ref{64random_long} and \ref{64points_long}. 
(The peak power for $L_z = 20$ in normalized units is 
$1.14 \cdot 10^{-6} \pm 9 \cdot 10^{-8}$ compared to 
$4.2 \cdot 10^{-7} \pm 3 \cdot 10^{-8}$ for $L_z = 60$.) 
Qualitatively, the findings are quite similar to those for a shorter vortex 
length; however, in the presence of point defects, compared to the results 
shown in Fig.~\ref{64points}, the narrowband power is lower.  
On the other hand, a similar intensity drop is not observed in the case of 
columnar defects, compare Fig.~\ref{64random_long} to Fig.~\ref{64random}.  
The additional flux line length and the lack of spatial correlation in the $z$ 
direction for point disorder further demonstrates the difference in pinning 
efficiency between point and columnar defects.  
For columnar pins the lengths of both the defect and the vortex span the height
of the sample just as they did for the shorter system; hence we observe a 
similar effect on the motion and the power spectrum.  
For point defects, as the length of the vortex is increased the effect of a 
single point defect on the longer vortex decreases as a whole resulting in a 
smaller power output: local fluctuations are averaged out.  
We note that naturally these distinctions are only observable in 
three-dimension simulations.

\subsection{Variable Pinning Efficiency}

In this section we examine the effect of varying pinning efficiency on the
vortex structure factor and velocity fluctuation power spectrum.
We shall study the cases of different point pin strenghts and linear defects of
fixed strength but varying length, thus interpolating between uncorrelated 
point disorder and correlated columnar defects.
We are particularly interested in whether point and columnar defects display 
different washboard harmonic signatures in the velocity power spectrum.

\subsubsection{Variable Point Pinning Strength}
\label{variablepinstrength}

\begin{figure}
\begin{center}
  \subfigure{\label{powergraph}
  \includegraphics[scale=.55]{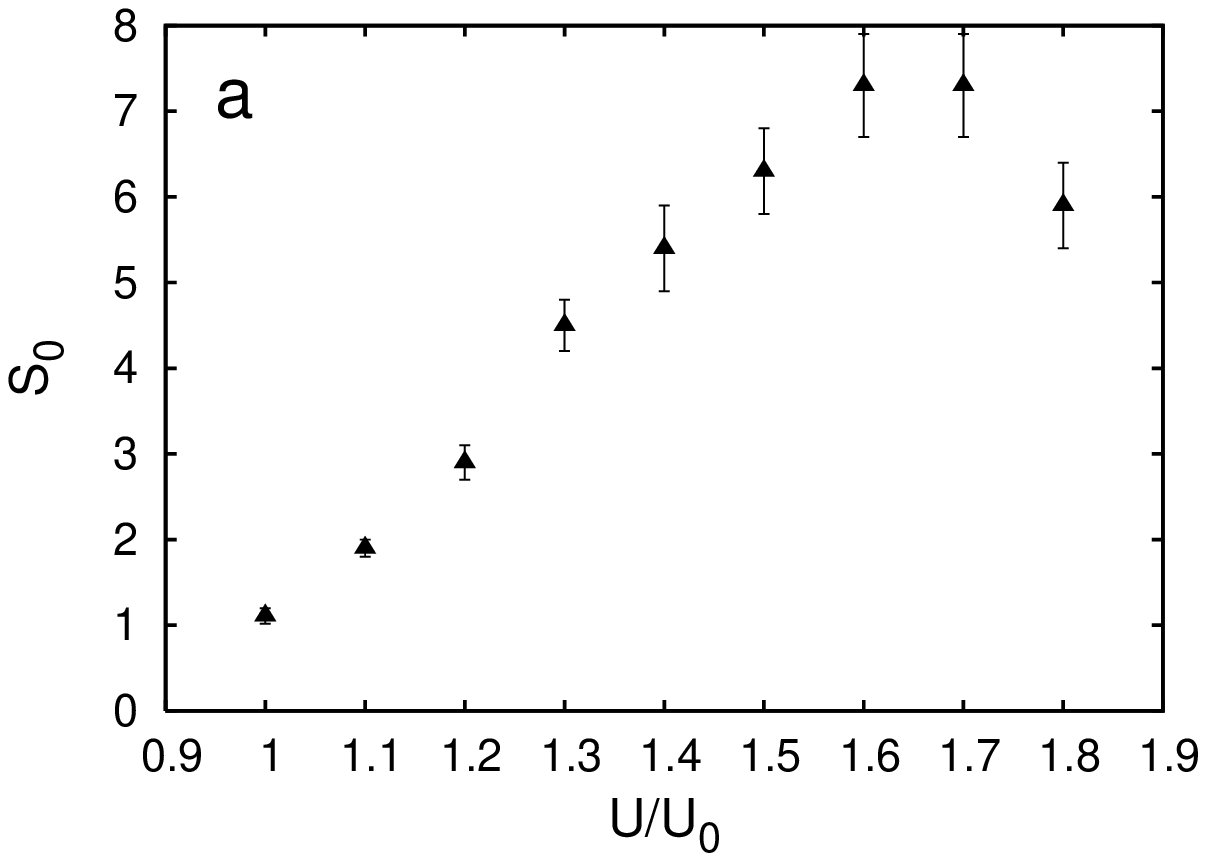}}
  \vskip -0.1 truecm
% \subfigure{\label{pinstrength1_1}
% \includegraphics[scale=.25]{./1_1_def_strength_xnoise.eps}}
  \subfigure{\label{pinstrength1_7}
  \includegraphics[scale=.25]{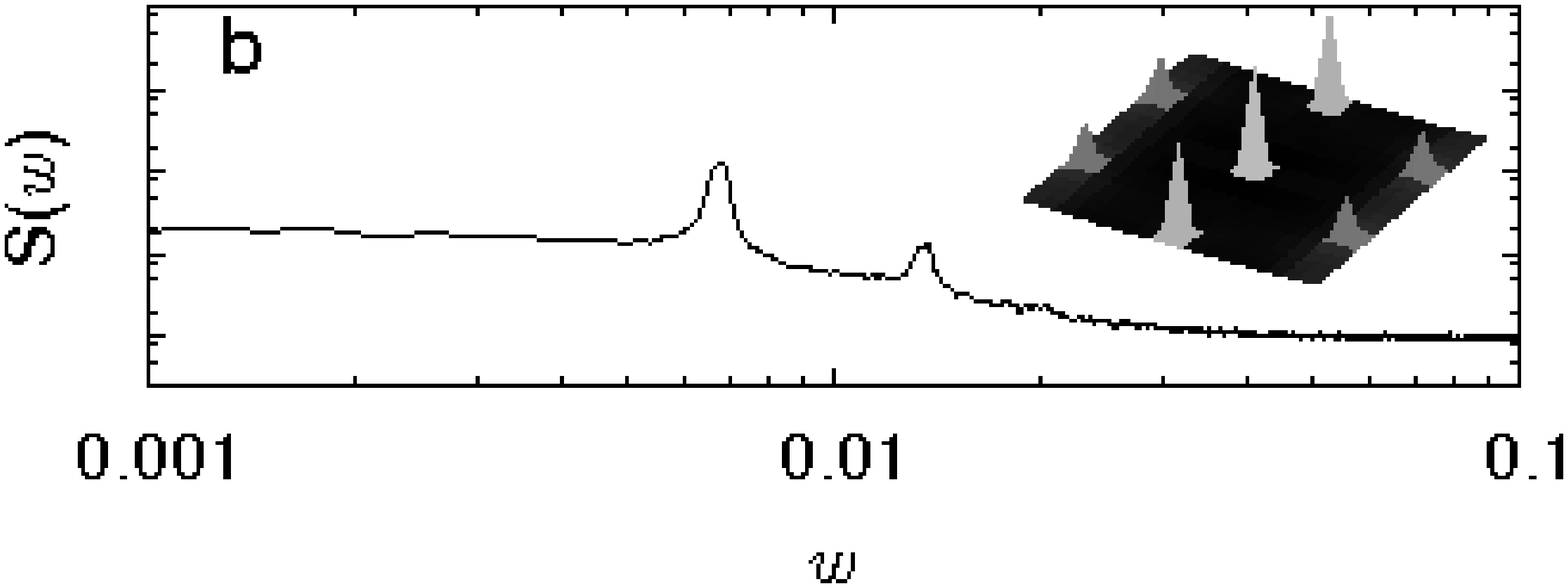}}
  \vskip -0.2 truecm
  \subfigure{\label{pinstrength1_9}
  \includegraphics[scale=.25]{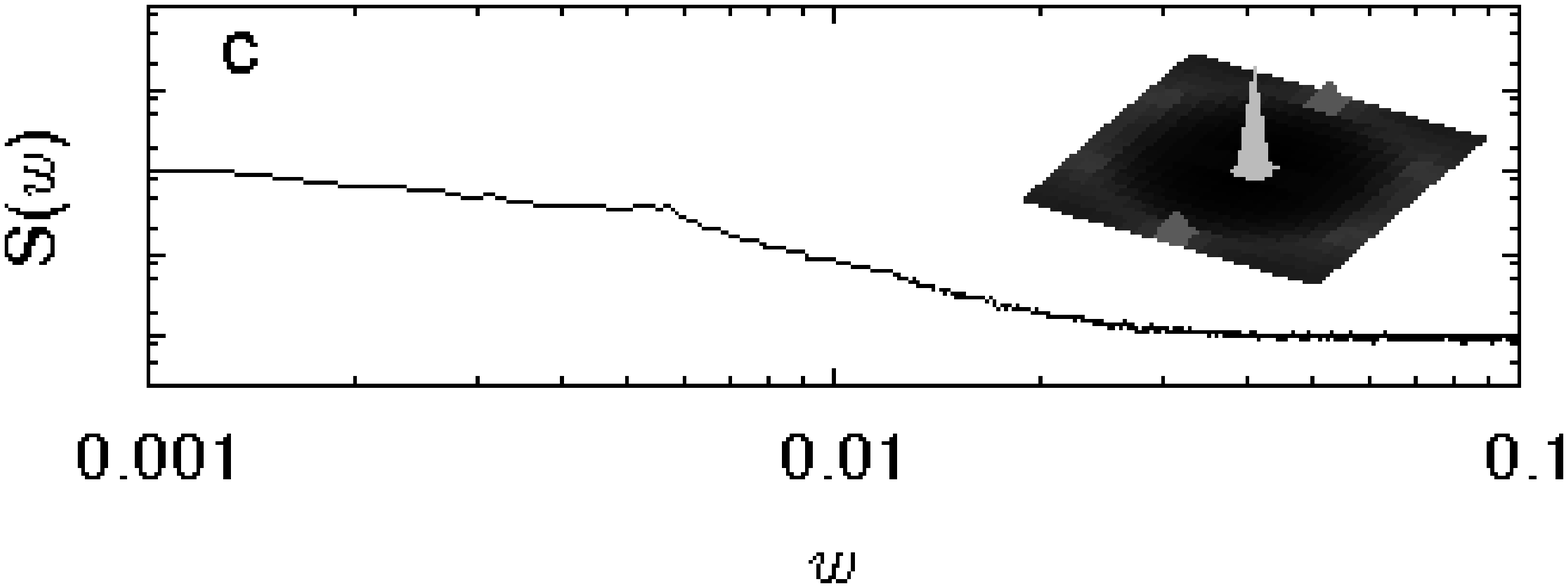}}
\end{center}
\vskip -0.3 truecm
\caption{(a) Velocity spectrum washboard peak power for increasing point 
  pinning strength $U$.  
  The vortex density is held constant at $\rho = 64 / A$.  
  As the pinning strength is increased the washboard signal grows to a maximum 
  value around $1.7 \, U_0$ (b) and then decreases as the system transitions to
  a broadband signal around $1.9 \, U_0$ (c).}
\label{power_v_pinstrength}
\end{figure}
To carry out this investigation simulation runs were performed at fixed flux 
density $\rho = 64 / A$ while the point pinning strength was increased from the
value $U_0$ used before up to $U = 1.9 \, U_0$.  
For each pinning strength results were obtained by averaging over random 
distributions of point defects.  
The number of pinning sites per run was held constant; only the pinning 
strength was changed.  
The power of the washboard peak is plotted versus pinning strength in 
Fig.~\ref{powergraph}.  
We note that the power reaches a maximum value at approximately $1.7 \, U_0$,
see Fig.~\ref{pinstrength1_7}. 
The increase in power is due to the stronger pinning causing larger velocity 
fluctuations in the vortex array.  
The width of the peak increases along with its power as indicated by the error 
bars in Fig.~\ref{powergraph}.  
The signal then decreases as the power output transitions to a broadband signal
coinciding with decreasing coherent vortex motion, Fig.~\ref{pinstrength1_9}.
 
\begin{table}
\caption{Intensity ratio of the three largest velocity power spectrum peaks for
  increasing point defect pinning strength, recorded as multiples of the 
  initial pinning strength $U_0$.  
  For each pinning strength the ratios of the intensities of the first and 
  second peak to the third are given for measurements taken in the drive ($x$)
  direction.  
  The number of runs over which the power spectral density plots were averaged
  is also included.}
\begin{center}
\begin{tabular}{|c|c|c|}
\hline $U / U_0$ & runs & ratio ($x$ direction) \\
\hline $1.1$ & $44$ & $1$ : $0.55\pm0.06$ : $0.150\pm0.020$ \\
\hline $1.3$ & $44$ & $1$ : $0.33\pm0.04$ : $0.045\pm0.008$ \\
\hline $1.5$ & $44$ & $1$ : $0.20\pm0.03$ : $0.012\pm0.004$ \\
\hline
\end{tabular}
\end{center}
\vskip -0.2 truecm
\label{pointpower_pinstren}
\end{table}
Washboard harmonic peak ratios for a few pinning strenghts are recorded in 
Table~\ref{pointpower_pinstren}.  
We compare these results to the columnar defect data in Table~\ref{columnpower}
and note that similar harmonic ratios are observed for the weaker columnar 
defects at a density of $\rho = 64 / A$ and the stronger point defects with
$U = 1.5 \, U_0$.  
These findings demonstrate that the harmonic ratios are not unique signatures 
of the material defect types compared in this study.  
We speculate that the ratios are rather dependent on the amount of deformation 
of the vortex lattice by the defects regardless of whether or not the pins 
extend along the length of the vortices.

\subsubsection{Variable Linear Defect Length}

\begin{figure}
\begin{center}
  \subfigure{\label{len1}\includegraphics[scale=.25]{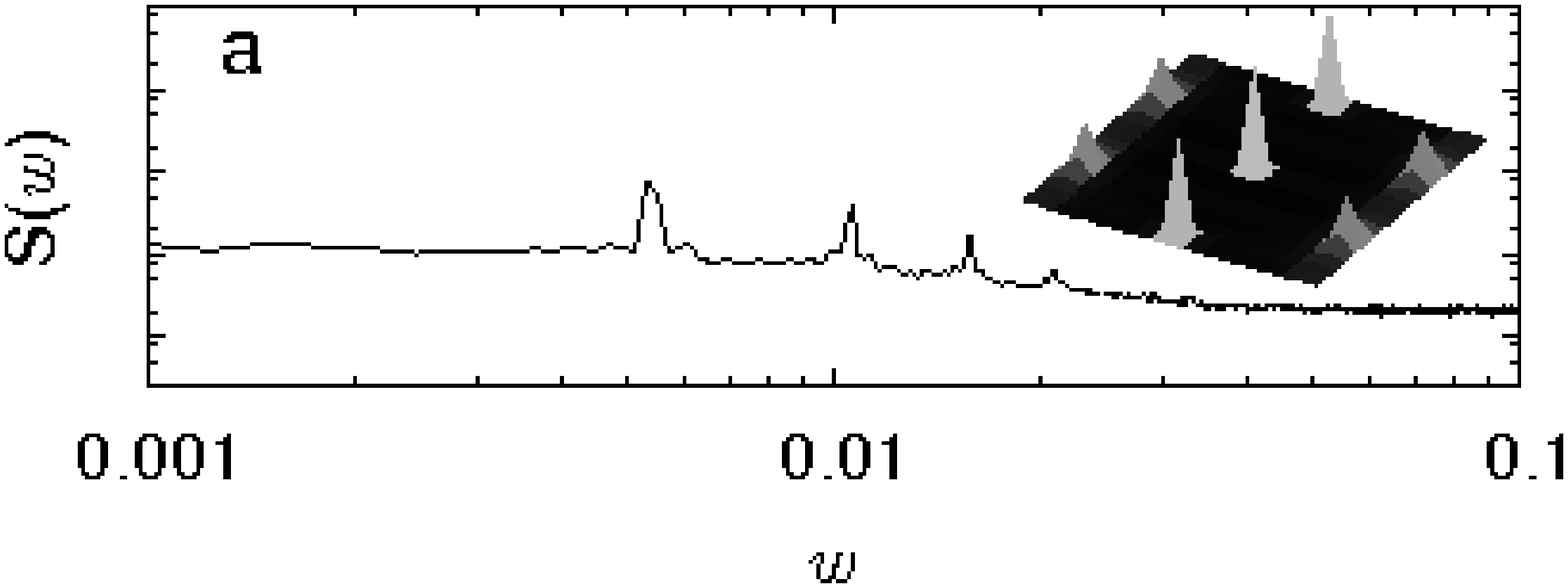}}
  \vskip -0.2 truecm
  \subfigure{\label{len3}\includegraphics[scale=.25]{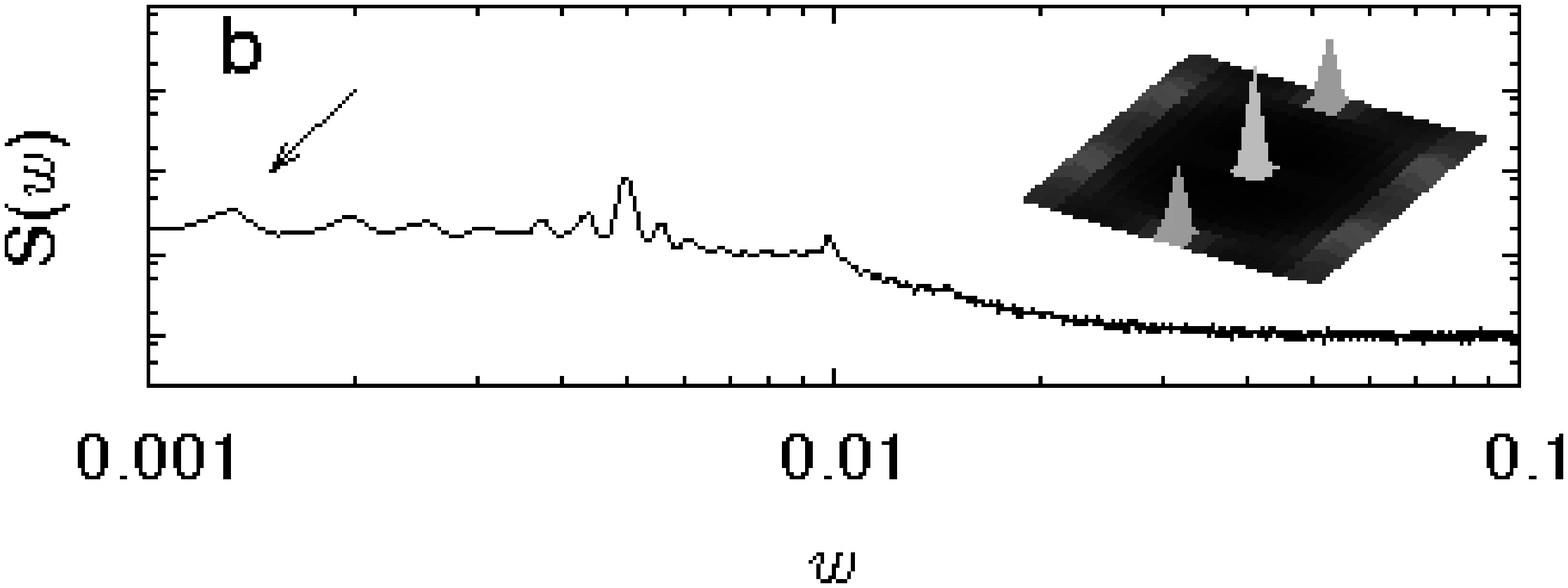}}
  \vskip -0.2 truecm
  \subfigure{\label{len5}\includegraphics[scale=.25]{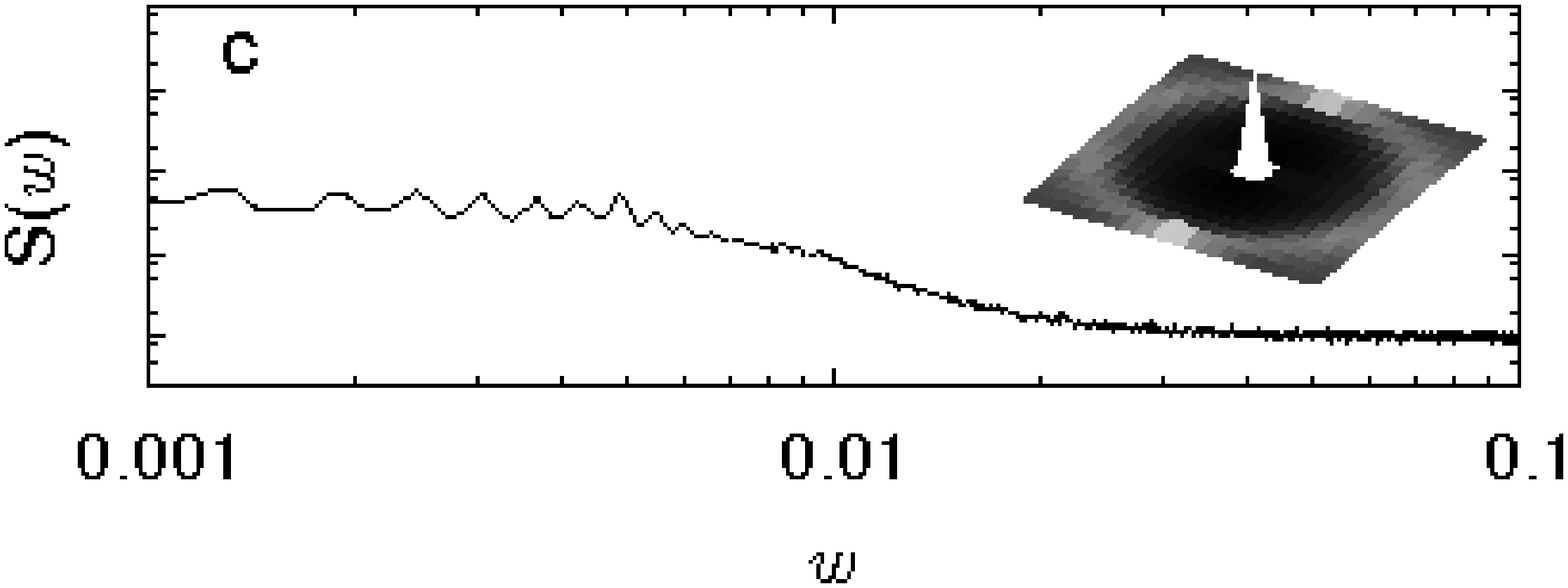}}
\end{center}
\vskip -0.3 truecm
\caption{Vortex velocity / voltage power spectra and diffraction plots for 
  increasing columnar defect lengths at a fixed vortex density 
  $\rho = 16 / A = 0.00062/b_0^2$ with (a) $l = 1$, (b) $l = 3$, and 
  (c) $l = 5$.
  The results reveal how the system evolves from an ordered to a disordered 
  configuration as the defect length increases.  
  The washboard signal decreases as the broadband noise grows.}
\label{def_length}
\end{figure}
To further compare the effects of point and columnar defects on the velocity 
power spectrum, we have investigated systems with varying columnar defect 
lengths at a constant vortex density $\rho = 16 / A$.  
Each set of results is obtained by averaging runs over random distributions of 
linear defects of a particular length $l$, with the total number of pinning 
sites held constant.  
Defect lengths vary from a single pinning site, i.e., random point defects, to
a length of $10$ contiguous pinning sites.  
(Recall that a columnar defect extending over $20$ pinning sites spans the 
entire system height $L_z = 20$.)  
The driving force remains at $f=0.04$.
As the defect length is increased the spatial order in the system decreases as
indicated by the structure factor peaks in Fig.~\ref{def_length}.
Peaks with wave vector components along the drive decrease in intensity at 
shorter defect lengths compared to those perpendicular to the drive.  
With increasing length of the linear pins (growing defect correlations) the 
peak intensities become diminished.  
Results for defect length $l = 1$, $3,$ and $5$ are shown in 
Fig.~\ref{def_length}.
By length $l = 10$ the washboard peak disappears entirely and is replaced by a 
broadband signal (not shown).
We note a second peak with a frequency corresponding to the system size also 
appears in the power spectrum $l = 3$.  
While the cause of this peak is unclear, one possible explanation is that it 
originates in some type of persistent deformation in the vortex lattice.  
With periodic boundary conditions this lattice deformation would repeatedly 
travel over the same pinning distribution resulting in a periodicity 
corresponding to the `time of flight' across the system. 
\begin{figure}
\begin{center}
 \subfigure{\label{def_1_1}\includegraphics[scale=.25]{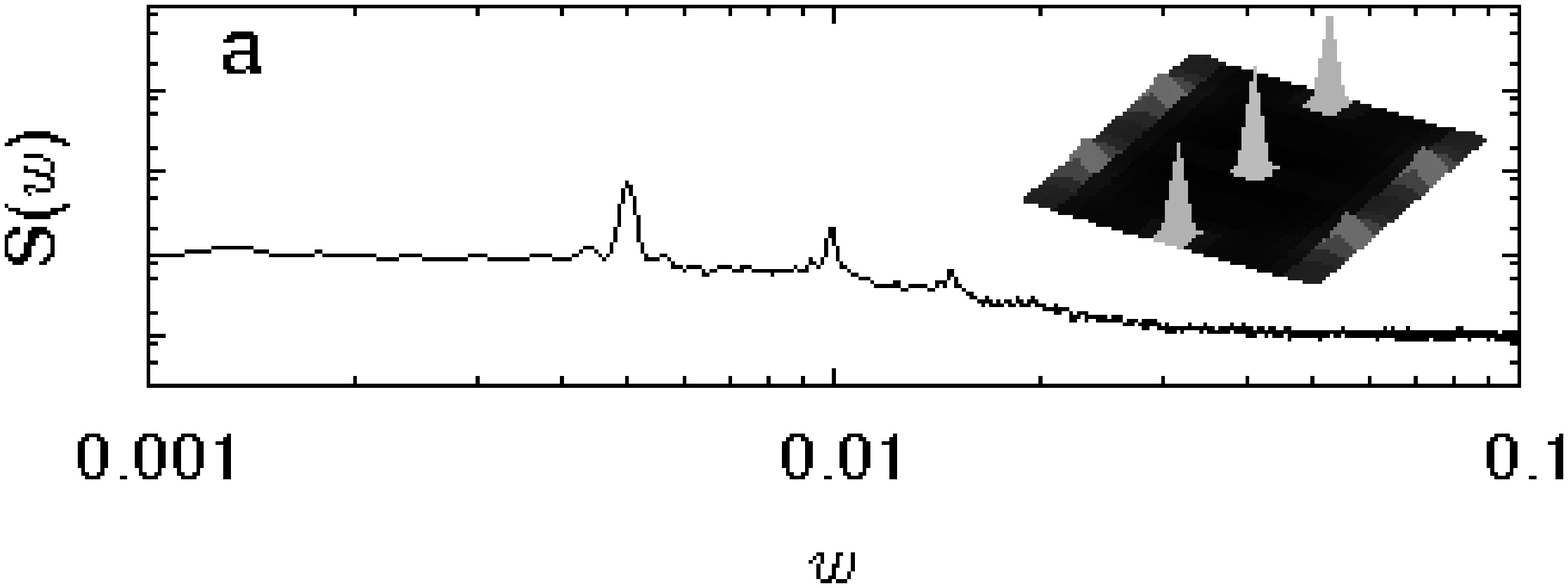}}
 \vskip -0.2 truecm
 \subfigure{\label{def_1_3}\includegraphics[scale=.25]{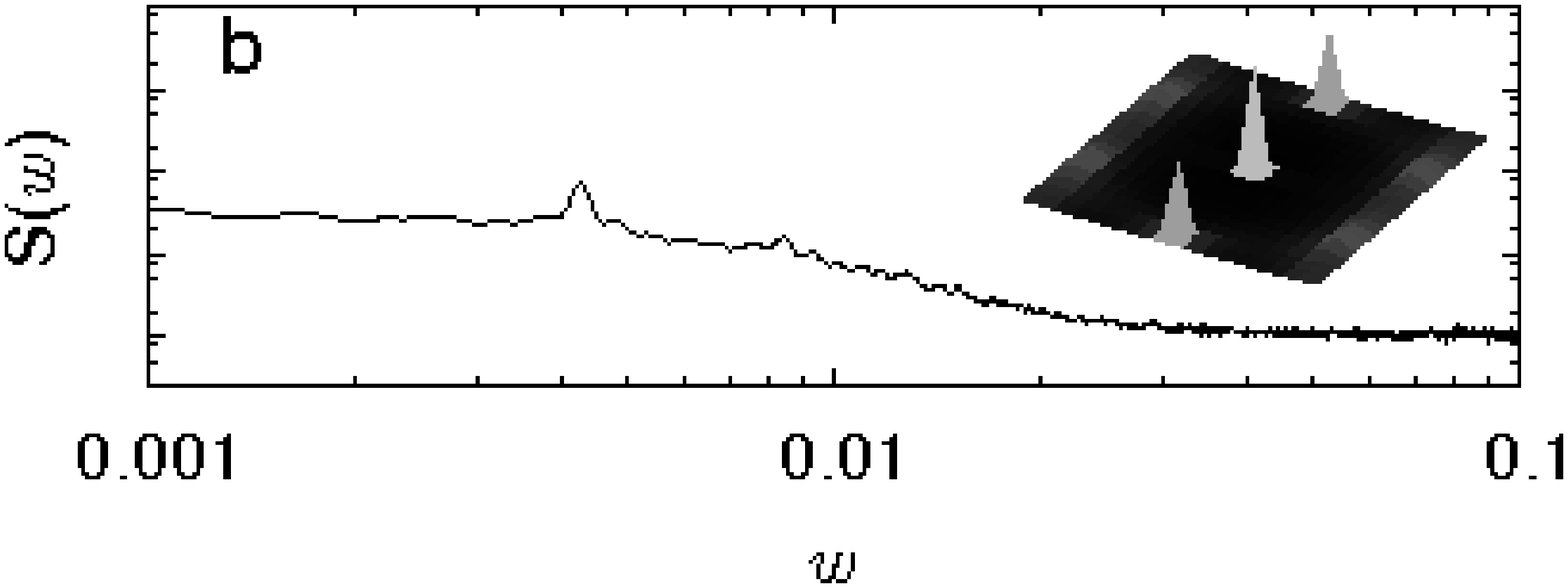}}
 \vskip -0.2 truecm
 \subfigure{\label{def_1_4}\includegraphics[scale=.25]{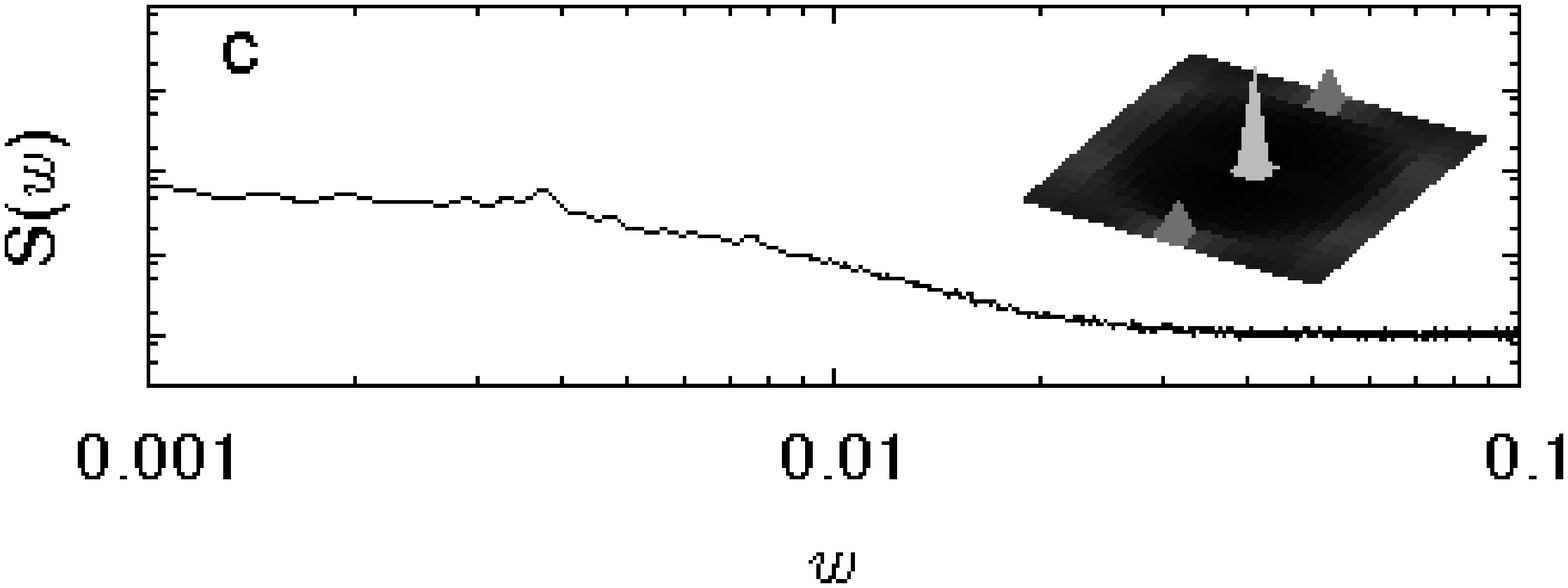}}
\end{center}
\vskip -0.3 truecm
\caption{Results for simulations with random point defects whose pinning 
  strength is increased from its original value $U_0$ in (a) to 
  (b) $U = 1.2 \, U_0$ and $U = 1.4 \, U_0$ at a low vortex density 
  $\rho = 16 / A = 0.00062/b_0^2$.} 
\label{def_strength}
\end{figure}
As a comparison runs were performed for various point defect pinning strengths
at the same low flux density $\rho = 16 / A$.  
The process was identical to the previous section.  
However, due to the lower density, when the depth of the pinning potential well
was increased to only $1.5$ times its original value $U_0$, the washboard peak 
completely disappeared.  
The results of these simulations are shown in Fig.~\ref{def_strength}.  
We observe that these results look quite similar to those from increasing the
linear defect lengths. 
 
\begin{figure}
\begin{center}
  \subfigure{\includegraphics[scale=.6]{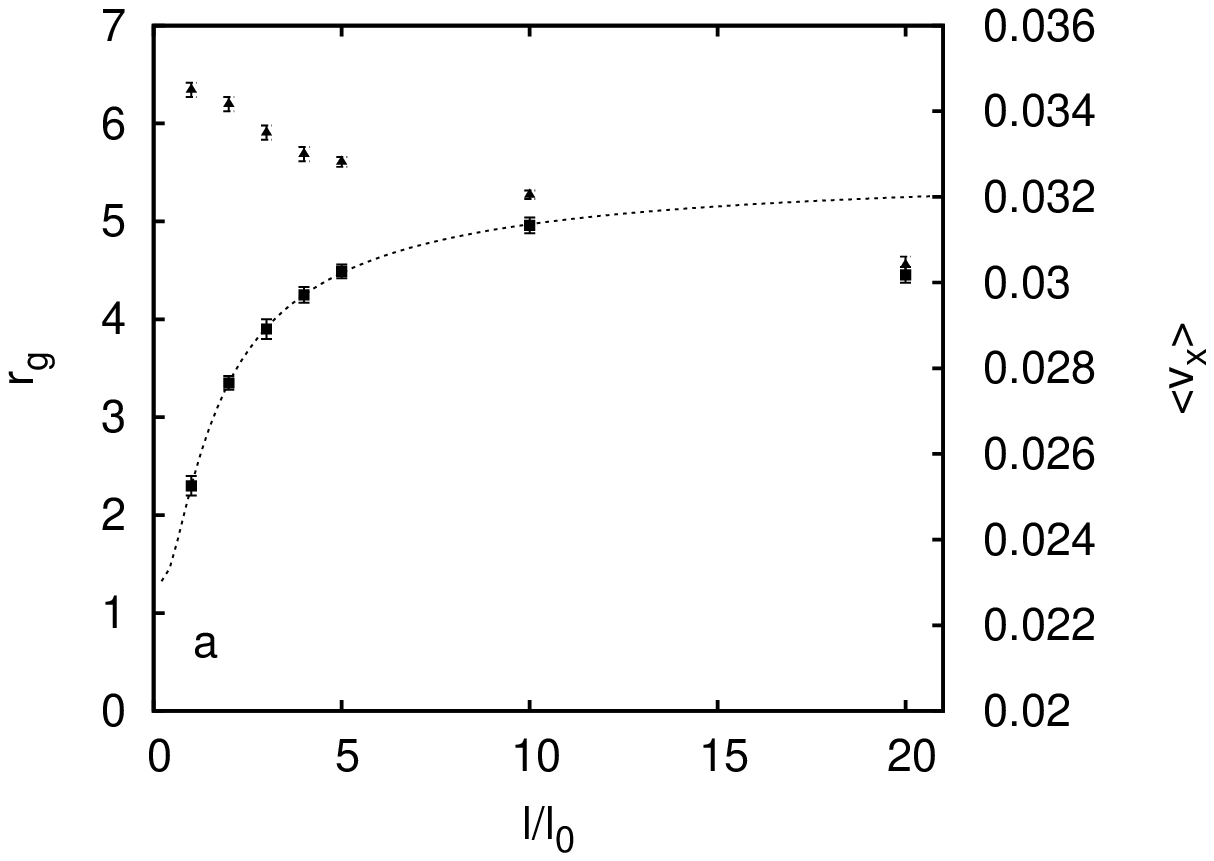}}
  \vskip -0.2 truecm
  \subfigure{\includegraphics[scale=.6]{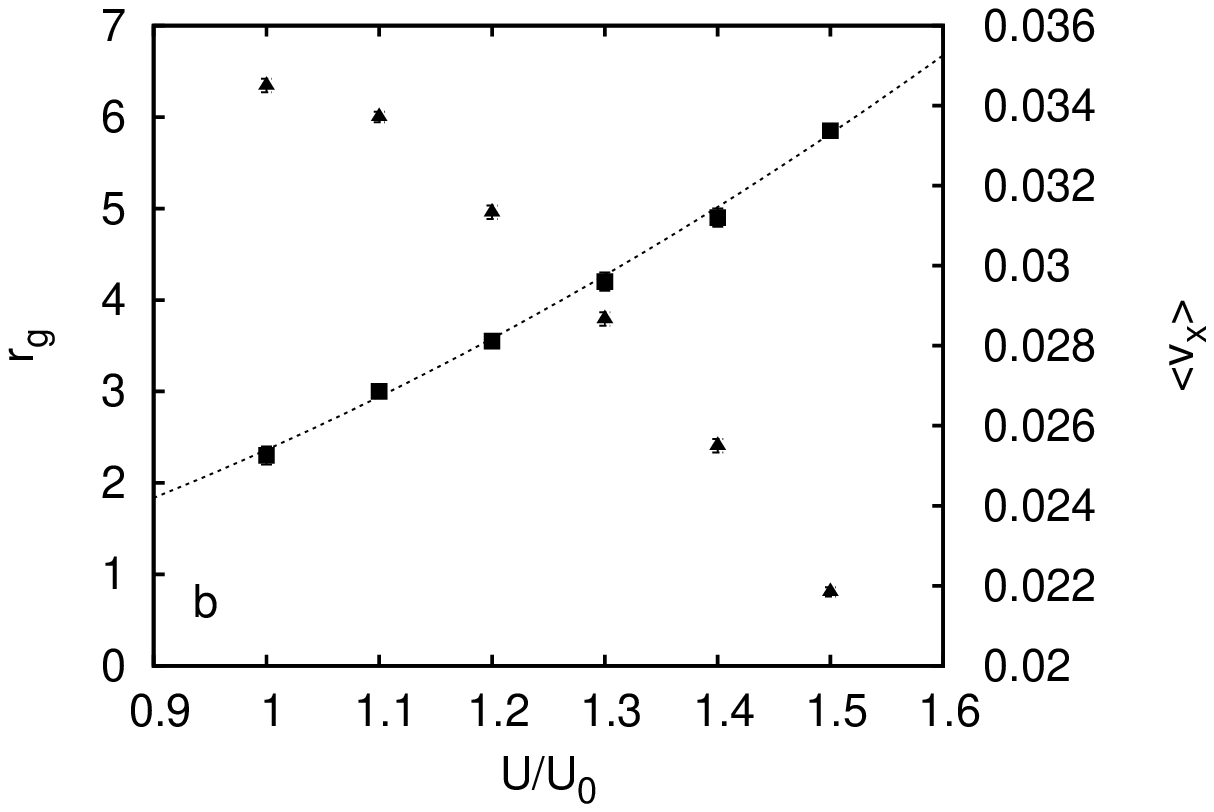}}
\end{center}
\vskip -0.6 truecm
\caption{Mean flux line radius of gyration and average velocity for 
  (a) increasing linear defect length and (b) enhanced point pinning strength.
  The radius of gyration data for increasing lengths initially follows an 
  exponential of the form $r_g \propto \exp(- l_0/l)$ while the data for 
  increasing pinning strength is best described by a quadratic fit. 
  $\blacktriangle$ - average velocity, $\blacksquare$ - radius of gyration 
  ($x$ component).}
\label{def_str_vs_col_len}
\end{figure}
While the above power spectra are qualitatively comparable, the growth of the 
mean vortex line radius of gyration turns out to be quantitatively different in
both situations, as shown in Fig.~\ref{def_str_vs_col_len}.  
The data for columnar disorder initially grows as an exponential function, 
$r_g \propto \exp(- l_0/l)$ where $l$ is the length of the columnar defects, 
while the data for point defects is best described by a quadratic fit 
$r_g \propto (U / U_0)^2$.  
While we cannot offer a quantitative theory for these trends, they can be
understood qualitatively by examining the average vortex velocity as these two 
pinning types are varied.  
For both cases as the `pinning effectiveness' is increased the velocity 
decreases and the radius of gyration increases.  
We interpret the increase in radius of gyration as the vortices being stretched
as they are simultaneously held by the defects and pulled by the externally 
applied drive.  
As the linear defect length is increased the velocity decreases, but by length 
$l = 20$ (the height of the system) the vortices are still moving.  
[Compare the velocity trends in Fig.~\ref{def_str_vs_col_len}(a) and (b).]
This indicates that the columnar defects cannot stop the vortices at this 
particular flux density and applied drive, and the `stretching' of the vortices
is limited leading to saturation of the radius of gyration.  
On the other hand, for the point pins the pinning strength is not limited 
leading to greater stretching.

\section{Summary and Conclusion}
\label{sec:Summary and Conclusion}

To summarize, a non-equilibrium Monte Carlo simulation code has been developed 
to study the effects of different types of pinning centers on the velocity 
power spectrum of vortices driven through a disordered medium in the 
non-equilibrium steady state \cite{TOM,DAS}.   
Specifically, we have investigated the evolution of the washboard signal as we
have increased the vortex / magnetic flux density in the presence of point and 
columnar defects.  
In order to achieve a more complete understanding of the velocity power 
spectra, we have also examined the corresponding two-dimensional vortex 
structure factor and the average flux line radius of gyration.

We have confirmed that our numerical model displays the appropriate physical 
behavior over a large section of parameter space.  
For instance, vortices arrange into a six-fold lattice when interacting with 
only weak material defects.  
In the presence of an applied current and sufficiently strong pinning sites, 
the vortex lattice reorients with a lattice vector along the drive direction.  
Furthermore, we have obtained the current-voltage characteristics and found 
these to be qualitatively similar to I-V curves obtained in experiments.  
In our simulation vortices depin above a critical applied force and gain 
velocity as the applied force is increased.  
Changes in the I-V curves due to different pinning site types and vortex 
densities occur as expected.

At high applied drive, our algorithm produces saturating I-V curves, an 
artifact that originates in a necessary limitation of the maximum allowed 
displacements. 
However, we have provided evidence that suggests this limitation does not 
adversely affect the associated power spectra and other physical quantities in 
the regime investigated here.

Expected physical behavior is also observed in the vortex structure factor.  
For both pointlike and linear extended defects, as the vortex density is 
increased spatial ordering is observed to increase in the diffraction plots.  
When interacting with columnar pins, the vortex system structure factor is 
found to evolve from that of a liquid, to a smectic, and finally a triangular 
lattice.  
For point defects only the regular triangular array is realized in the 
parameter space studied here.  
We remark that the structure factor plots with point defects at low vortex 
densities appear qualitatively similar to those for columnar defects at higher 
flux densities.

Velocity fluctuation or voltage noise power spectra measured parallel to the 
drive have been obtained for various vortex densities in the presence of both 
columnar and point defects.  
A narrowband signal is observed over a large vortex density interval, with the 
peak coinciding with the washboard frequency.  
The power spectrum has also been measured perpendicular to the drive, and a 
signal at the washboard frequency is observed there as well.  
Harmonics have been detected at multiples of the washboard frequency for both 
types of disorder.   
For columnar defects the ratios of the power of the first and second harmonic 
to the third turn out to be larger than for point defects.  
However, as the density is increased, the ratio decreases for columnar defects.
By varying the pinning strength of the point defects at a constant vortex 
density we have obtained similar harmonic ratios to that of columnar defects.  
This indicates that the harmonic ratios are not unique indicators of the type 
of material defects present in the sample.  
Rather, we think that the harmonic ratios should be a function of the degree of
deformation of the vortex lattice.
We remark that the detailed features of the flux flow power spectrum are also
influenced by the configuration of the leads above the sample surface 
\cite{CLEM}.
In real experimental setups, these effects may mask the noise signatures 
observed in our simulations. 

In order to investigate the shape of the fluctuating vortex lines we have 
determined the average radius of gyration in the presence of point and columnar
defects in the $x$ and $y$ directions.  
The general (and expected) trend is for the radius of gyration to decrease as 
the density of the vortices is increased.  
Our results for the gyration radius also reveal that transverse fluctuations do
not appear to play a large role in the thermal flux line wandering for the 
range of parameters investigated here:
The transverse component in fact rarely exceeds the radius of a pinning site 
except for the lowest vortex densities studied.

In addition, we have measured these various observables as the columnar defect 
length, i.e., the degree of correlation in the disorder was varied, and 
compared the results to the effects of varying just the pinning strength of 
point defects.  
Changing the defect character from point-like to columnar shows similar results
to increasing point defect pinning strength in both the diffraction and power 
spectral density plots.  
Different effects on the vortices through adjusting these two distinct pinning 
mechanisms become apparent upon comparing the growth of the radius of gyration 
in the direction along the drive.  
While the results seem to correspond to physical intuition, a precise theory 
explaining the quantitative growth of the radius of gyration in either 
situation is currently lacking and will have to be developed.

\begin{acknowledgement}
This work was in part supported through the U.S. National Science Foundation, 
Division of Materials Research, grants NSF DMR-0075725 and 0308548, and 
through the Bank of America Jeffress Memorial Trust, research grant J-594.  
Some of the data shown were obtained from simulations run on Virginia Tech's 
Anantham cluster.  
We gratefully acknowledge helpful discussions with I. Georgiev, 
T. Klongcheongsan, E. Lyman, M. Pleimling, G. Pruessner, B. Schmittmann, 
S. Teitel, and R.K.P. Zia.
\end{acknowledgement}


\begin{thebibliography}{00}

\bibitem{BLAT}
For a detailed review, see: G. Blatter, M. V. Feigel'man, V. B. Geshkenbein, 
A. I. Larkin, and V. M. Vinokur, Rev. Mod. Phys. {\bf 66} 1125 (1994).

\bibitem{CLEM}
J. R. Clem, Phys. Rep. {\bf 75}, 1 (1981).

\bibitem{BRANDT}
E. H. Brandt, Phys. Rev. Lett. {\bf 50}, 1599 (1983).

\bibitem{MATS1}
T. Matsuda, K. Harada, H. Kasai, O. Kamimura, and A. Tonomura,
Science {\bf 271}, 1393 (1996).

\bibitem{JENS1}
H. J. Jensen, A. Brass, and A. J. Berlinsky, 
Phys. Rev. Lett. {\bf 60}, 1676 (1988).

\bibitem{VEST}
A. Vestergren and M. Wallin, Phys. Rev. B {\bf 69}, 144522 (2004).

\bibitem{YARIN}
U. Yaron, P. L. Gammel, D. A. Huse, R. N. Kleiman, C. S. Oglesby, E. Bucher, 
B. Batlogg, D. J. Bishop, K. Mortensen, K. Clausen, C. A. Bolle, and 
F. De~La~Cruz, Phys. Rev. Lett. {\bf 73}, 2748 (1994).

\bibitem{KOSH}
A. E. Koshelev and V. M. Vinokur, Phys. Rev. Lett. {\bf 73}, 3580 (1994).

\bibitem{DOUS1}
T. Giamarchi and P. Le~Doussal, Phys. Rev. Lett. {\bf 76}, 3408 (1996); 
P. Le~Doussal and T. Giamarchi, Phys. Rev. B {\bf 57}, 11356 (1998).

\bibitem{GOTCHA}
V. Gotcheva, A.T.J. Wang, and S. Teitel, Phys. Rev. Lett. 92, 247005 (2004);
V. Gotcheva, Y. Wang, A. T. J. Wang, and S. Teitel, 
Phys. Rev. B {\bf 72}, 064505 (2005).

\bibitem{CHEN}
Q.-H. Chen and X. Hu, Phys. Rev. Lett. {\bf 90}, 117005 (2003).

\bibitem{GRUNER}
G. Gr\"{u}ner, Rev. Mod. Phys. {\bf 60}, 1129 (1988);
{\em Density Waves in Solids} (Addison-Wesley, Reading, MA 1994).

\bibitem{CIVALE}
L. Civale, A. D. Marwick, T. K. Worthington, M. A. Kirk, J.R. Thompson, 
L. Krusin-Elbaum, Y. Sun, J. R. Clem, and F. Holtzberg, 
Phys. Rev. Lett. {\bf 67}, 648 (1991).

\bibitem{NEL1}
D. R. Nelson and V. M. Vinokur, 
Phys. Rev. Lett. {\bf 68}, 2398 (1992); Phys. Rev. B {\bf 48}, 13060 (1993).

\bibitem{OLIVE}
E. Olive, J. C. Soret, P. Le Doussal, and T. Giamarchi, 
Phys. Rev. Lett. {\bf 91}, 037005 (2003).

\bibitem{FIORY}
A. T. Fiory, Phys. Rev. Lett. {\bf 27}, 501 (1971).

\bibitem{HARRIS}
J. M. Harris, N. P. Ong, R. Gagnon, and L. Taillefer,
Phys. Rev. Lett. {\bf 74}, 3684 (1995).

\bibitem{TROY}
A. M. Troyanovski, J. Aarts, and P. H. Kes, 
Nature (London) {\bf 399}, 665 (1999).

\bibitem{TOGAWA}
Y. Togawa, R. Abiru, K. Iwaya, H. Kitano, and A. Maeda,
Phys. Rev. Lett. {\bf 85}, 3716 (2000).

\bibitem{OLSON1}
C. J. Olson, C. Reichhardt, and F. Nori, 
Phys. Rev. Lett. {\bf 81}, 3757 (1998).

\bibitem{KOLTON1}
A. B. Kolton, D. Dom\'nguez, and N. Gr\/onbech-Jensen, 
Phys. Rev. B {\bf 65}, 184508 (2002).

\bibitem{FANG}
H. Fangohr, S. J. Cox, and P. A. J. de Groot, 
Phys. Rev. B {\bf 64}, 064505 (2001).

\bibitem{TOM}
T. J. Bullard, Ph. D. Thesis, 
Virginia Polytechnic Institute and State University (2005).

\bibitem{DAS}
J. Das, T. J. Bullard, and U. C. T\"auber, Physica A {\bf 318}, 48 (2003).

\bibitem{ROSSO}
A. Rosso and W. Krauth,	Phys. Rev. B {\bf 65}, 012202 (2001).

\bibitem{SEN}
P. Sen, N. Trivedi, and D. M. Ceperley, Phys. Rev. Lett. {\bf 86}, 4092 (2001).

\bibitem{fnote1}
We observe that with a severe cut-off the system becomes locked into a 
particular spatial configuration, as each line is effectively caged within the 
deep potential energy minima formed by linear superposition of its neighbors' 
step-function potentials. % created by the cut-off. 

\bibitem{KATZ}
S. Katz, J. L. Lebowitz, and H. Spohn, Phys. Rev. B {\bf 28} 1655 (1983).

\bibitem{OLSON2}
C. J. Olson, C. Reichhardt, and F. Nori, 
Phys. Rev. Lett. {\bf 80}, 2197 (1998).

\bibitem{KLEIN}
W. H. Kleiner, L. M. Roth, and S. H. Autler, 
Phys. Rev. {\bf 133}, A1226 (1964).

\bibitem{SCHMID}
A. Schmid and W. Hauger, J. Low Temp. Phys. {\bf 11}, 667 (1973).

\bibitem{ANDO}
Y. Ando, N. Motohira, K. Kitazawa, J.I. Takeya, and S. Akita, 
Phys. Rev. Lett. {\bf 67}, 2737 (1991).

\bibitem{AMMOR}
L. Ammor, B. Pignon, and A. Ruyter, Phys. Rev. B {\bf 69}, 134508 (2004).

\bibitem{QIANG}
Q. Li, H. J. Wiesmann, M. Suenaga, L. Motowidlow, and P. Haldar,
Phys. Rev. B {\bf 50}, 4256 (1994).

\bibitem{OLSON3}
C. J. Olson, G. T. Zim\'anyi, A. B. Kolton, and N. Gr\/onbech-Jensen,
Phys. Rev. Lett. {\bf 85}, 5416 (2000).

\bibitem{KOLTON2}
A. B. Kolton, D. Dom\'inguez, C. J. Olson, and N. Gr\/onbech-Jensen,
Phys. Rev. B {\bf 62}, R14 657 (2000).

\end{thebibliography}
\end{document}